\documentclass[a4paper,twocolumn,11pt]{quantumarticle}
 \pdfoutput=1
\usepackage[utf8]{inputenc}
\usepackage[english]{babel}
\usepackage[T1]{fontenc}
\usepackage{amsmath}
\usepackage{hyperref}
\usepackage{tikz}
\usepackage{lipsum}
\usepackage{amsmath,amssymb,amsfonts} 
\usepackage{amsthm}                   
\usepackage{physics}                  
\usepackage{graphicx}                 
\usepackage{booktabs,array}           
\usepackage{multirow}                 
\usepackage{dsfont}                   
\usepackage{tikz}                     
\usetikzlibrary{quantikz}             

\begin{document}

\title{Quantum Simulation of Dynamical Transition Rates in Open Quantum Systems}

\author{Robson Christie}
\affiliation{School of Mathematics and Physics, University of Portsmouth, PO1 3FX, United Kingdom}

\author{Kyunghyun Baek}
\affiliation{Institute for Convergence Research and Education in Advanced Technology, Yonsei University, Seoul 03722, Republic of Korea}

\author{Jeongho Bang}
\affiliation{Institute for Convergence Research and Education in Advanced Technology, Yonsei University, Seoul 03722, Republic of Korea}

\author{Jaewoo Joo}
\affiliation{School of Mathematics and Physics, University of Portsmouth, PO1 3FX, United Kingdom}
\orcid{0000-0003-0491-980X}
\email{jaewoo.joo@port.ac.uk (corresponding author)}

\begin{abstract}
Estimating transition rates in open quantum systems is hampered by computing-resource demands that grow rapidly with system size. We present a quantum-simulation framework that enables efficient estimation by recasting the transition rate, given as the time derivative of an equilibrium correlation function, into a set of independently measurable contributions. Each contribution term is evaluated as the expectation value of a parameter-tuned quantum process, thereby circumventing explicit Lindbladian numerics. We validate our method on a spin-$\frac{1}{2}$ decoherence model using an IBM quantum processor. Further, we apply the method to the Caldeira–Leggett model of quantum Brownian motion as a realistic and practically relevant setting and reaffirm the theoretical soundness and practical implementability. These results provide evidence that quantum simulation can deliver substantial computational advantages in studying open-system kinetics for quantum chemistry on an intermediate-scale quantum computer. 
\end{abstract}

\maketitle

\section{Introduction}
\begin{figure}[t]
\centering
\includegraphics[width=0.48\textwidth,trim=0cm 0.5cm 0cm 0cm]{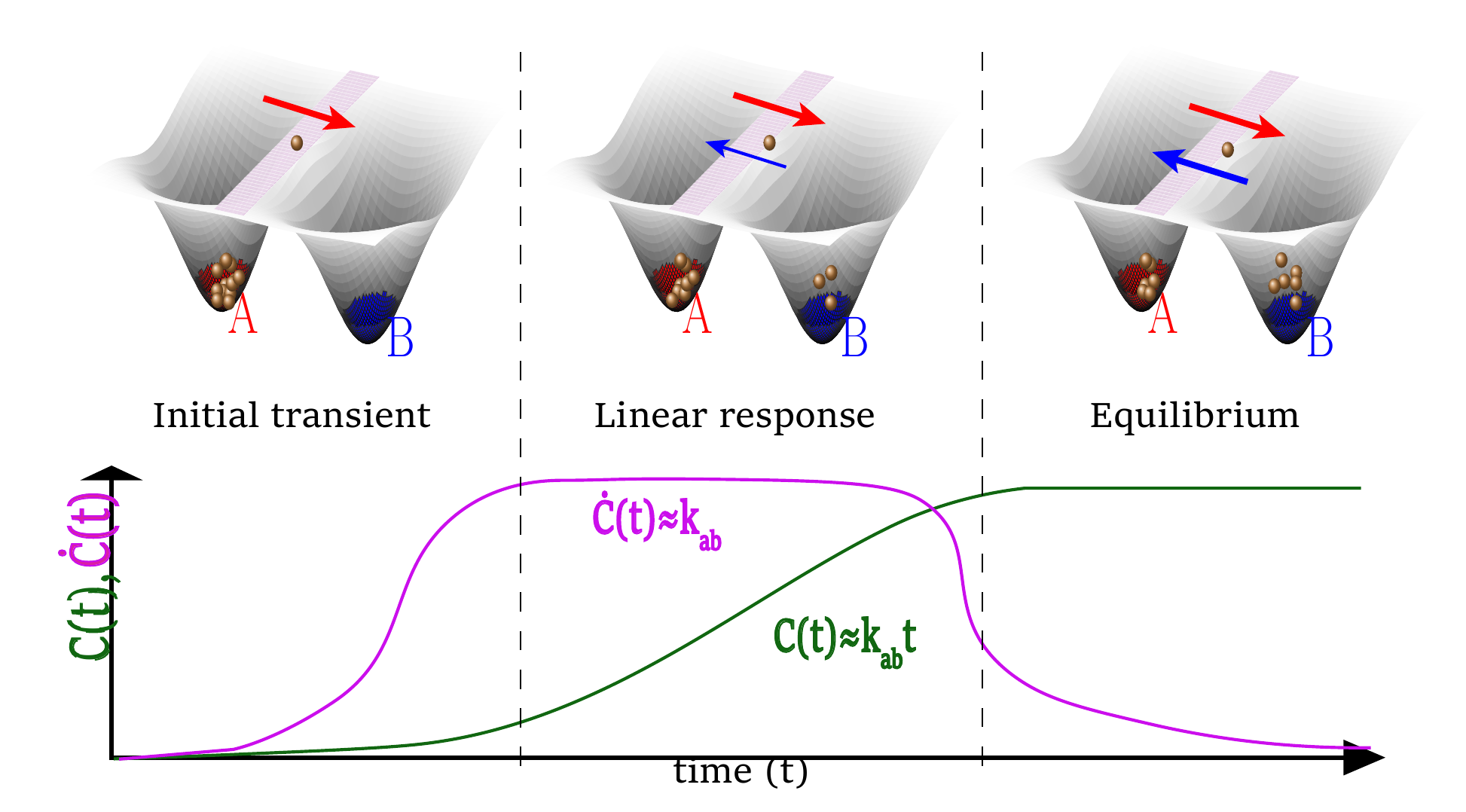}
\caption{Metastable-transition dynamics: $C(t)$ (green) grows approximately linearly after an intrabasin transient, while the transition rate $\dot C(t)$ (violet) exhibits a plateau $\dot C(t)\!\approx\!k_{AB}$ for $\tau_{\rm intra}\!\ll\!t\!\ll\!\tau_{\rm eq}$ before decaying to zero at equilibrium ($\tau_{\rm intra}$: intrabasin relaxation time and $\tau_{\rm eq}$: global equilibrium time).}
\label{fig:cartoon1}
\end{figure}

Estimating transition rates are central to chemistry and statistical physics, underpinning barrier crossing, diffusion, and relaxation \cite{Schlosshauer2007}.
For metastable open systems governed by Markovian master equations, two dynamical objects suffice to characterize the rates: the equilibrium correlation $C(t)$ and its time derivative $\dot C(t)$. 
In the flux-side correlation formalism, after a short intrabasin transient starting from $A$ to $B$, the dynamics enter a linear-response regime in which $\dot C(t)$ approaches a plateau shown in Fig.~\ref{fig:cartoon1}. 
The interbasin rate $k_{AB}$ is then read off as $k_{AB}\simeq \dot C(t)$ for $\tau_{\rm intra}\ll t\ll \tau_{\rm eq}$ before $\dot C(t)\to0$ at global equilibrium. 
Despite considerable advancements in classical numerics, estimating $k_{AB}$ remains challenging: Lindblad propagation scales poorly with Hilbert-space dimension; plateau extraction from long-time correlators is sample-hungry; and complex environments introduce noise and memory effects \cite{Hsieh2013,Hele2017}. 

Quantum simulation has rapidly matured into a powerful tool for exploring complex quantum dynamics beyond state-of-the-art classical numerics, with notable demonstrations across quantum chemistry \cite{mcardle2020quantum,guo2024experimental,chan2023grid,su2021fault,bauer2020quantum}, many-body physics \cite{busnaina2024quantum,chertkov2023characterizing,cygorek2022simulation}, quantum field theory~\cite{bauer2023quantum,davoudi2021toward,schafer2020tools}, and cosmology~\cite{antonini2020cosmology,barata2024probing,viermann2022quantum}. 
In the near term, noisy intermediate-scale quantum (NISQ) devices motivate methods tailored to limited qubit counts, shallow depths, and hardware noise~\cite{preskill2018quantum,arute2019quantum,chen2023complexity}. 
Simulation of open-system dynamics using quantum master-equation primitives particularly incorporates environmental effects through non-unitary channels, avoiding explicit bath simulation and improving scalability for complex open dynamics~\cite{samach2022lindblad,david2023digital,lee2023evaluating,liu2024simulation,guimaraes2024optimized,cleve2016efficient,joo2023commutation}, 
with the potential to establish a quantum-enhanced toolkit for real-world applications~\cite{campaioli2024quantum}. 

Motivated by these challenges, we develop a hardware-ready framework that estimates transition rates by reducing the time derivative of the quantum time-correlation function to a finite set of independently measurable expectation values such as
\begin{eqnarray}
k_{AB}\approx \frac{\sum_{j}\langle \hat O_j\rangle_{\Phi}}{\langle \hat O^\star\rangle}.
\label{eq:k_sch}
\end{eqnarray}
Each expectation contribution is obtained by executing a parameter-tuned completely positive trace-preserving process (CPTP) \(\Phi\) and measuring a single observable \(\hat O_j\) (with normalization by \(\langle \hat O^{\star}\rangle\)). 
This derivative-to-expectation reduction dispenses with explicit Lindbladian integration and long-time trajectory averaging: all quantities are single-time, shallow-depth, and parallel across terms. 
We validate the workflow on an IBM quantum processor (IBMQ) using a spin-$1/2$ decoherence model and find close agreement with theory across a range of dissipation strengths. 
We then apply the scheme to the Caldeira-Leggett model of quantum Brownian motion \cite{CLmodel83,Breuer2002} where system-bath structure matters and demonstrate its stability and implementability over experimentally relevant parameter ranges. 
Taken together, these results introduce a distinct route to the investigation of kinetics on quantum hardware and indicate a new credible path toward practical quantum advantages. 

The reduced dynamics of an open quantum system coupled to an environment are described by a CPTP map. 
When bath correlations decay on short timescales compared with the system’s evolution, the Markovian limit is realized. Then, the dynamical maps form a semigroup generated by the Lindblad master equation with its system $\hat{\rho}_S$ ~\cite{gorini1976,lindblad1976generators,mccauley2020accurate}
\begin{eqnarray}
\hspace{-1cm} && \frac{d}{dt}\hat{\rho}_S
= \mathcal{L} \left( \hat{\rho}_S (t) \right) 
= -\frac{i}{\hbar}[\hat{H},\hat{\rho}_S] \nonumber \\
\hspace{-1cm} &&~~~~~~~~~ + \frac{1}{\hbar}\sum_{k}\!\left(\hat{L}_k \hat{\rho}_S \hat{L}_k^{\dagger}
- \frac{1}{2} \{ \hat{\rho}_S , \hat{L}_k^{\dagger} \hat{L}_k \} \right),
\label{eq:lindblad}
\end{eqnarray}
where $[\cdot,\cdot]$ and $\{\cdot,\cdot\}$ denote the commutator and anticommutator, $\hat H$ generates the unitary part, and the set of $\hat L_k$ encodes environmental effects (rate coefficients absorbed into $\hat L_k$).  
For instance, Hermitian $\hat L_k$ describes pure dephasing, whereas non-Hermitian $\hat L_k$ captures energy exchange such as dissipation and thermalization~\cite{petruccione,manzano2020short}. 

To characterize transitions between coarse-grained regions $A$ and $B$ of phase space, we introduce the one-dimensional (1D) projectors 
\begin{eqnarray}
\hat\theta_{A}=\!\int_{x\in A}\! \mathrm dx\, \ket{x}\bra{x},~~~~~
\hat\theta_{B}=\!\int_{x\in B}\! \mathrm dx\, \ket{x}\bra{x},
\end{eqnarray}
with position eigenstates $\ket{x}$. 
By $\langle \hat O\rangle_{\rm eq}\equiv{\rm Tr}(\hat\rho_{\rm eq}\hat O)$ for the equilibrium average with $\mathcal L(\hat\rho_{\rm eq})=0$, the normalized correlation $C(t)$ and its time derivative $\dot C(t)$ are given by
\begin{eqnarray}
C(t)=\frac{\langle \{\hat\theta_A(0),\hat\theta_B(t)\}\rangle_{\rm eq}}{2\,\langle \hat\theta_A(0)\rangle_{\rm eq}}, \\
\dot C(t)=\frac{\langle \{\hat\theta_A(0),\dot{\hat\theta}_B(t)\} \rangle_{\rm eq}}{2\,\langle \hat\theta_A(0)\rangle_{\rm eq}},
\label{eq:CandCdot-real}
\end{eqnarray}
where $\hat\theta_B(t)=e^{{\mathcal L}^\dagger t}(\hat\theta_B (0))$ and $\dot{\hat\theta}_B(t)={\mathcal L}^\dagger \!\left(\hat\theta_B(t)\right)$ in the Heisenberg picture. 
Equivalently, one evolves the state in time and keeps $\hat\theta_B$ as s fixed projector to obtain $C(t)$ and $\dot C(t)$ in the Schrödinger picture such as 
\begin{eqnarray}
&& C(t) = \frac{ \tr \Big( \hat\theta_B\,e^{\mathcal L t} \bigl( \bigl\{ \hat\rho_{\mathrm{eq}}, \hat\theta_A \bigr\} \bigr) \Big)}{2\langle\hat\theta_A \rangle_{\mathrm{eq}}}, \label{eq:rate-Sch001} \\
&& \dot C(t) = \frac{ \tr \Big( \hat\theta_B \,{\mathcal L} \left(e^{{\mathcal L} t} \bigl( \bigl\{ \hat \rho_{\mathrm{eq}}, \hat\theta_A \bigr\} \bigr) \right) \Big)}{2\langle\hat\theta_A \rangle_{\mathrm{eq}}},~~
\label{eq:rate-Sch}
\end{eqnarray}
where $\hat\theta_{A/B} (0) \equiv \hat\theta_{A/B}$. The detailed derivation of Eqs.~(\ref{eq:rate-Sch001}) and (\ref{eq:rate-Sch}) is provided in the Appendix. 

\section{Parameter-tunable quantum process} 
\begin{figure}[t]
\centering
\scalebox{0.85}{%
\begin{quantikz}
  \lstick{$\hat c_{+}$} & \gate{\hat R(\chi)} & \ctrl{2} & \ctrl{1} & \ctrl{2} & \ctrl{1} & \gate{\hat H_d} & \meter{} \\
  \lstick{$ \hat\theta_B$}     & \qw                 & \qw      & \gate{\hat N} & \qw    & \swap{1} & \qw & \qw \\
  \lstick{$\hat\rho_{\rm eq}$} & \qw & \gate{\hat\theta_{A,\epsilon}} & \gate{\mathrm e^{\mathcal L t}} & \gate{\hat M} & \targX{} & \qw & \qw \\
  \slice{$\hat\Phi_0$}  & \slice{$\hat\Phi_1$} & \slice{$\hat\Phi_2$} & \slice{$\hat\Phi_3$} & \slice{$\hat\Phi_4$} & \slice{$\hat\Phi_5$} & \slice{$\hat\Phi_6$} &
\end{quantikz}}
\caption{Parameter-tunable quantum process for estimating $C(t)$ and $\dot{C}(t)$. $\hat c_{+}$ is a pure state while $ \hat\theta_B$ and $\hat\rho_{\rm eq}$ are given as mixed states. 
The control qubit is phase-shifted by $\hat{R}(\chi)$ and read out after $\hat{H}_d$. $\hat{\theta}_{A,\epsilon}$ is the small-angle surrogate in Eq.~\eqref{eq:thetaA_approx}. $e^{{\mathcal L} t}$ is the CW-Lindblad propagator, and $\hat{N}$ and $\hat{M}$ are Hamiltonian/jump/anticommutator channels. The states $\hat\Phi$ in each step are explained in the Appendix.}
\label{fig:QCirc}
\end{figure}

\subsection{Correlation derivative circuit} 
We developed a parameter-tunable quantum circuit shown in Fig.~\ref{fig:QCirc} and realize the derivative-to-expectation reduction to single-time expectation values using a control qubit initialized in $\hat c_+=\ket{+}\bra{+}$ for $\ket{+} = (\ket{0} + \ket{1})/\sqrt{2}$ and two system registers. 
The first register is prepared in the projector state $\hat\theta_B$ (normally expressed by a mixed state), and the second in the equilibrium state $\hat\rho_{\rm eq}$ with ${\mathcal L}(\hat\rho_{\rm eq})=0$. 
On the control qubit, we apply a phase gate $\hat R(\chi)$, and we conditionally enact operators $\hat\theta_{A,\epsilon}$, $\hat N$, $\hat M$, and a controlled-SWAP (C-SWAP) gate between the two registers ($\epsilon$: angle parameter in a projector form). 
The open-system evolution block $e^{{\mathcal L} t}$ is implemented via the repeated-interaction Cleve–Wang (CW) scheme (see details in the Appendix) \cite{cleve2016efficient}. The detailed method of the C-SWAP gate with mixed-state inputs is also explained in the Appendix. After a Hadamard gate $\hat H_d$, measuring {\em only} the control qubit in the $Z$-basis yields the scalar 
\begin{eqnarray}
\mathcal E~ (\chi; \hat N,\hat M;t, \epsilon) = \Tr(\hat{\sigma}_z^{c} \hat \Phi_6) = \langle \hat\sigma_z^{c}\rangle , \label{eq:general_E}
\end{eqnarray}
from which contribution terms for $C(t)$ and $\dot C(t)$ are assembled (Pauli matrices: $\hat{\sigma}_{x,y,z}$). 

From the set of single-time expectation values in Eqs.~(\ref{eq:rate-Sch001}) and (\ref{eq:rate-Sch}), we obtain the linear-combination estimators
\begin{eqnarray}
C(t) &=& \frac{\mathcal E_C}{2\,\mathcal E_D}, \label{eq:dCdtSchro-1} \\
\dot C(t) &=& \frac{\mathcal E_{H1}+\mathcal E_{H2}+\mathcal E_J - (\mathcal E_{AC1}+\mathcal E_{AC2})/2}{2\hbar\,\mathcal E_D},~~~\nonumber \\
\label{eq:dCdtSchro}
\end{eqnarray}
where it shows $\mathcal E_{H1/H2}$ for Hamiltonian parts, $\mathcal E_J$ for the jump term, $\mathcal E_{AC1/2}$ for the anticommutator contributions in Eq.~\eqref{eq:lindblad}. In addition, for $\chi = 0$, $\mathcal E_C$ and  $\mathcal E_D$ are given by $\hat N = \hat M = \hat{\mathds{1}}$ and $\hat\theta_B = \hat N = \hat M = \mathrm e^{\mathcal L t} = \hat{\mathds{1}}$, respectively (details given in the next subsection). 
Each $\mathcal E$ is estimated from a {\em single} circuitry configuration, which implies that no explicit Lindbladian integrals, long-time fits, or two-time correlator reconstructions are required. 

When $\hat N$ or $\hat M$ is Hermitian but not unitary, we build the operator through the small-angle unitary surrogate
\begin{eqnarray}
\hat{\mathcal O}_\epsilon =\frac{e^{i \epsilon \hat{\mathcal O}}-e^{-i \epsilon \hat{\mathcal O}}}{2i \epsilon},
\label{eq:Ohat_approx}
\end{eqnarray}
whose expectation value admits the rapidly convergent expansion
\begin{eqnarray}
\langle \hat{\mathcal O}_\epsilon \rangle_{\rho}
= \langle \hat{\mathcal O} \rangle_{\rho}
- \frac{\epsilon^{2}}{3!} \langle \hat{\mathcal O}^{3} \rangle_{\rho}
+ \frac{\epsilon^{4}}{5!} \langle \hat{\mathcal O}^{5} \rangle_{\rho} - \cdots,
\label{eq:expansion}
\end{eqnarray}
and equals $\langle \hat{\mathcal O} \rangle_{\rho}$ to leading order. 
In particular, applying the same construction for the projector $\hat{\theta}_A = (\hat\theta_A)^k$ ($k$: integer) yields
\begin{equation}
\hat{\theta}_{A,\epsilon}=\frac{\sin\epsilon}{\epsilon}\,\hat\theta_A\,.
\label{eq:thetaA_approx}
\end{equation}
In addition, this prefactor contributaion is cancelled between its numerator and denominator in Eqs.~(\ref{eq:rate-Sch001}) and (\ref{eq:rate-Sch}), so both $C(t)$ and $\dot C(t)$ become independent of $\epsilon$ for $\hat\theta_A$. 

\begin{figure}[t]
\centering
\scalebox{0.95}{%
\begin{quantikz}
    \lstick{$\hat{c}_{+}$}  & \ctrl{2} & \qw & \ctrl{1} & \gate{\hat{H}_d} & \meter{} \\
    \lstick{$\hat{\theta}_B$} & \qw & \qw & \swap{1} & \qw & \qw \\
    \lstick{$\hat{\rho}_{eq}$} & \gate{\hat{\theta}_{A,\epsilon}} & \gate{e^{\mathcal{L} t}} & \targX{} & \qw & \qw \\
    \slice{ } & \slice{ } & \slice{ } & \slice{ } & \slice{ } &
  \end{quantikz}}
\caption{Simplified circuit of Fig.~\ref{fig:QCirc} for evaluating $\mathcal{E}_C$. \label{fig:QCircSupp}}
\end{figure}

\begin{table}[t]
\centering
\setlength{\tabcolsep}{0.09in}
\renewcommand{\arraystretch}{1.4}
\begin{tabular}{|c||c|c|c|c|c|}
\hline
${\mathcal E}$ & ${\mathcal E}_{H1}$ & ${\mathcal E}_{H2}$ & ${\mathcal E}_J$ & ${\mathcal E}_{AC1}$ & ${\mathcal E}_{AC2}$ \\
\hline
$\chi$ & $-\frac{\pi}{2}$ & $\frac{\pi}{2}$ & $0$ & $0$ & $0$ \\
\hline
$\hat{N}$   & $\hat{\mathds{1}}$ & $\hat H$ & $\hat L^{\dagger}_k$ & $\hat{\mathds{1}}$ & $\hat L^{\dagger}_k \,\hat L_k$ \\
\hline
$\hat{M}$  & $\hat H$ & $\hat{\mathds{1}}$ & $\hat L_k$ & $\hat L^{\dagger}_k \,\hat L_k$ & $\hat{\mathds{1}}$ \\
\hline
\end{tabular}
\caption{The set of parameter settings for Fig.~\ref{fig:QCirc}. Each column yields one expectation value from the control-qubit measurements ${\mathcal E} =\langle \hat{\sigma}_z^{c} \rangle$ to assemble Eq.~\eqref{eq:dCdtSchro}.}
\label{tab:TermTable}
\end{table}

\subsection{Expectation value set for $C(t)$ and $\dot C(t)$} 
The denominator \emph{$\mathcal{E}_{D}$} in Eqs.~(\ref{eq:dCdtSchro-1}) and (\ref{eq:dCdtSchro}) is simply calculated by the Hadamard test with two mixed-state inputs representing $\hat{\theta}_A$ and $ \hat{\rho}_{eq}$.
The correlation function numerator \emph{$\mathcal{E}_{C}$} in Eq.~(\ref{eq:dCdtSchro-1}) is computed with Fig.~\ref{fig:QCircSupp}, which is a simplified version of Fig.~\ref{fig:QCirc} with $\chi=0$, $\hat{N}=\hat{\mathds{1}}$, and $\hat{M}=\hat{\mathds{1}}$ 
\begin{eqnarray}
       \mathcal{E}_{C}=\Tr(\hat{\theta}_B \left( e^{ \mathcal{L}t} \left( \left\{\hat{\rho}_{eq} , \hat{\theta}_A \right\}\right)\right)).~
\end{eqnarray}
 
The correlation derivative numerator in Eq.~\eqref{eq:dCdtSchro} is given by the sum of five input combinations into the circuit in Fig.~\ref{fig:QCirc}.
For the Hamiltonian contributions (\emph{$\mathcal{E}_{H1}$} and \emph{$\mathcal{E}_{H2}$}), we first have $\chi=-\pi/2$, $\hat{N}=\hat{\mathds{1}}$, and $\hat{M}=\hat{H}$ and obtain 
\begin{eqnarray} 
&& \mathcal{E}_{H1} = \frac{i}{2}\bigg( \Tr(  \hat{\theta}_B  e^{\mathcal{L}t}\left( \hat{\rho}_{eq} \hat{\theta}_A\right) \hat H) \nonumber \\
&&~~~~~~~~~~~~ - \Tr(\hat{\theta}_B \hat H e^{\mathcal{L}t}\left(\hat{\theta}_A \hat{\rho}_{eq}\right))\bigg), ~~~\label{eq:HtermS11}
\end{eqnarray}
and choose $\chi=\pi/2$, $\hat{N}=\hat H$, and $\hat{M}=\hat{\mathds{1}}$ for
\begin{eqnarray}
&& \mathcal{E}_{H2} = \frac{ i}{2} \bigg( \Tr(\hat{\theta}_B  e^{\mathcal{L}t}\left(\hat{\theta}_A\hat{\rho}_{eq} \right)\hat H)  \nonumber \\
&&~~~~~~~~~~~~ - \Tr(  \hat{\theta}_B \hat H e^{\mathcal{L}t}\left( \hat{\rho}_{eq} \hat{\theta}_A\right) ) \bigg).  ~~~ \label{eq:HtermS2}
\end{eqnarray}
Combining Eqs.~\eqref{eq:HtermS11} and \eqref{eq:HtermS2}, we have 
\begin{equation} 
\mathcal{E}_{H1}+ \mathcal{E}_{H2}=-\frac{i}{2} \Tr(  \hat{\theta}_B\left[ \hat H ,e^{\mathcal{L}t}\left( \{\hat{\rho}_{eq}, \hat{\theta}_A\}\right)\right] ).
\end{equation}
For simplicity, with the assumption of single Lindblad operator $\hat L$, we set $\chi=0$, $\hat{N}=\hat L^{\dagger}$, and $\hat{M}=\hat L$ to estimate for the jump term  
\begin{eqnarray} \label{eq:Jterm}
&& \mathcal{E}_{J}=\frac{1}{2}\bigg(\Tr(  \hat{\theta}_B \hat L  e^{\mathcal{L}t}\left(\{ \hat{\rho}_{eq} ,\hat{\theta}_A\}\right) \hat L^{\dagger})\bigg).~~~~~
\end{eqnarray}

The anti-commutator contribution consists of two expectation value measurements. First with $\chi=0$, $\hat{N}=\hat{\mathds{1}}$ and $\hat{M}=\hat L^{\dagger} \hat L$, we estimate
\begin{eqnarray}
&& \mathcal{E}_{AC1} = \frac{1}{2} \bigg(  \Tr(  \hat{\theta}_B  e^{\mathcal{L}t}\left( \hat{\rho}_{eq} \hat{\theta}_A\right) \hat L^{\dagger} \hat L) \nonumber \\
&&~~~~~~~~~~~~~~ +  \Tr(\hat{\theta}_B \hat L^{\dagger} \hat L  e^{\mathcal{L}t}\left(\hat{\theta}_A\hat{\rho}_{eq}\right))\bigg), ~~~~~ \label{eq:ACtermS1}
\end{eqnarray}
and with $\chi=0$, $\hat{N}=\hat L^{\dagger} \hat L$ and $\hat{M}=\hat{\mathds{1}}$
\begin{eqnarray} 
&& \mathcal{E}_{AC2} = \frac{1}{2}\bigg( \Tr(  \hat{\theta}_B  \hat L^{\dagger} \hat L  e^{\mathcal{L}t}\left( \hat{\rho}_{eq} \hat{\theta}_A\right) )  \nonumber \\
&& ~~~~~~~~~~~~~~+ \Tr(\hat{\theta}_B   e^{\mathcal{L}t}\left(\hat{\theta}_A\hat{\rho}_{eq}\right)\hat L^{\dagger} \hat L)\bigg). ~~~~~ \label{eq:ACtermS2}
\end{eqnarray}
Again, combining Eqs.~\eqref{eq:ACtermS1} and \eqref{eq:ACtermS2}, we have
\begin{equation} \label{eq:ACtermS}
\mathcal{E}_{AC1}+ \mathcal{E}_{AC2}=\frac{1}{2}\Tr(  \hat{\theta}_B \left\{ e^{\mathcal{L}t}\left( \left\{\hat{\rho}_{eq}, \hat{\theta}_A\right\}\right), \hat L^{\dagger} \hat L \right\} )
\end{equation}
Therefore, with the combination of all the five terms in Eq.~(\ref{eq:dCdtSchro}), we finally evaluate the quantum transition rate $\dot C (t)$ with the parameter-tunable quantum circuits.

\begin{table}[t]
\centering
\begin{tabular}{|c|c|c|}
\hline
$\hat{\rho}_{eq}$ & $\hat{\theta}_{A}$ & $\mathcal{E}$ \\ \hline
$\ket{0}\bra{0}$ & $\hat{\mathds{1}}$ & $\frac{1}{2} \left( 1 - e^{- \gamma_0\, t /\hbar} f (t) \right) $ \\ \hline
$\ket{0}\bra{0}$ & $\hat \sigma_z$ & $\frac{1}{2} \left( 1 - e^{- \gamma_0 \, t /\hbar} f (t) \right) $ \\ \hline
$\ket{1}\bra{1}$ & $\hat{\mathds{1}}$ & $\frac{1}{2} \left( 1 + e^{- \gamma_0 \, t /\hbar} f (t) \right)$ \\ \hline
$\ket{1}\bra{1}$ & $\hat \sigma_z$& $-\frac{1}{2} \left( 1 + e^{- \gamma_0 \, t /\hbar} f (t) \right)$ \\ \hline
\end{tabular}
\caption{Analytic expectation value expressions contributing to the correlation function in Eq.~\eqref{eq:Ct001} for the decohering spin-1/2 model with parameters in Eq.~\eqref{eq:spin12_met2} $ \left( f (t) =  \cosh\left(\omega t \right) + \frac{\gamma_0 }{\hbar \omega}\sinh\left(\omega t \right) \right)$.
\label{tab:CTermTable}}
\end{table}

\begin{figure*}[t]
\centering
\begin{minipage}[b]{0.44\linewidth}
  \includegraphics[width=\textwidth,trim=0cm 0cm 1cm 0cm]{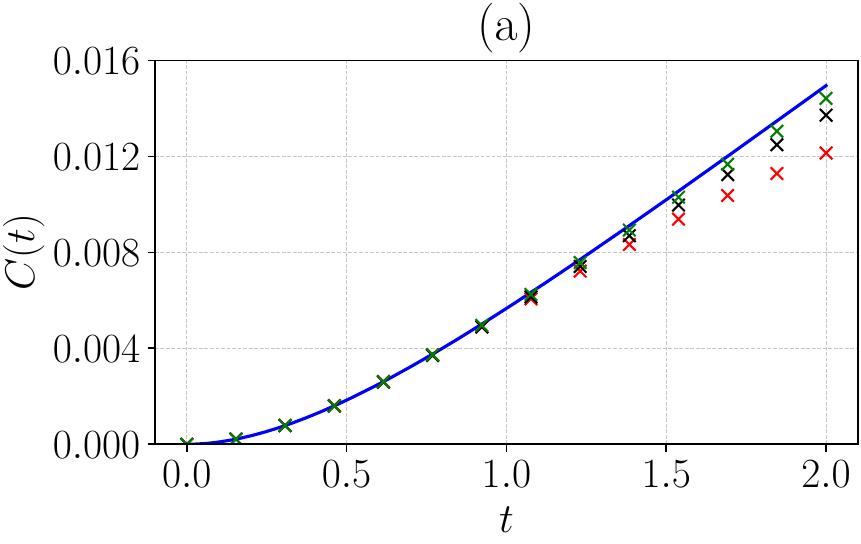}
\end{minipage}\hfill
\begin{minipage}[b]{0.45\linewidth}
  \includegraphics[width=\textwidth,trim=1cm 0cm 0cm 0cm]{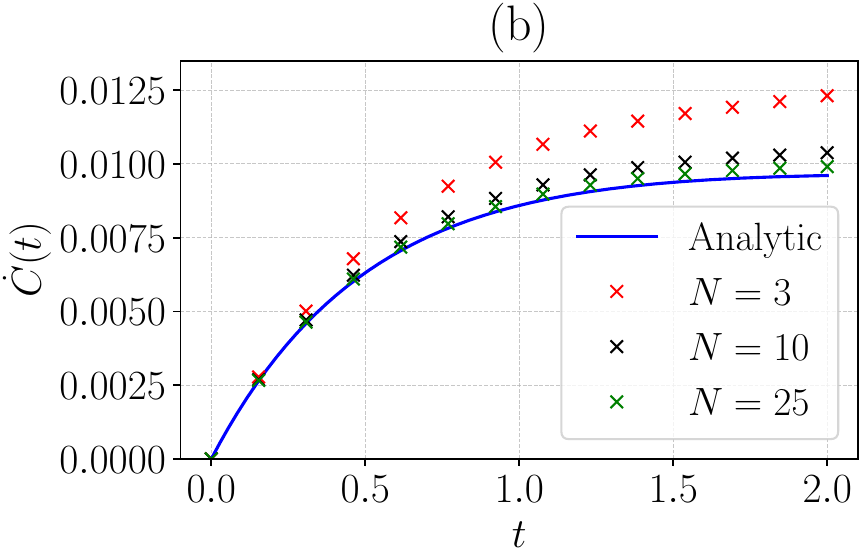}
\end{minipage}
\caption{Spin-$\frac{1}{2}$ testbed. (a) $C(t)$ and (b) $\dot C(t)$ versus $t$ for $\mu=0.1$, $\gamma_0=1$, $\hbar=1$: analytics (blue, solid) vs CW-Lindblad QuTiP emulation at $N= 3$, 10, and 25 time steps.}
\label{fig:spin-half01}
\end{figure*}

\section{Motivative example: spin-$\frac{1}{2}$ model}
\subsection{Analytical and numerical results for decohering spin-$\frac{1}{2}$ transitions}
A simple open quantum system to study the transition rate is a single spin-1/2 system with a magnetic field and continuous measurements.
We benchmark the estimator on an analytically solvable spin-$\frac{1}{2}$ system driven by a transverse field and continuously monitored in two spin states $\ket{0}$ and $\ket{1}$. 
This system is described by a Lindblad equation of the same form in Eq.~\eqref{eq:lindblad} with Hamiltonian and Lindbladian operators such as
\begin{equation} \label{eq:spin12_met2}
    \hat H = \mu \hat \sigma_y \quad \text{and} \quad \hat L =\sqrt{\gamma_0} \, \hat \sigma_z\,.
\end{equation}

In the dephasing-dominated regime $\mu\ll\gamma_0 \,$, the pointer states $\ket{0}$ and $\ket{1}$ are metastable ($\gamma_0$: decoherence rate). 
Choosing the projectors of $\hat{\theta}_A=\ket{0}\bra{0}$ and $\hat{\theta}_B=\ket{1}\bra{1}$, the stationary state is known as $\hat{\rho}_{\rm eq}=\frac12(\ket{0}\bra{0}+\ket{1}\bra{1})$. 
We can analytically solve these Lindblad dynamics for the correlation function and its rate expressions in
\begin{eqnarray}
\hspace{-1cm} && C(t) = \frac{1}{2}\!\left[1-e^{-{\gamma_0 t}/{\hbar}}\!\left(\cosh\omega t+\frac{\gamma_0}{\hbar\omega}\sinh\omega t\right)\right], \nonumber \\
\hspace{-1cm} && \label{eq:Ct001} \\
\hspace{-1cm} && \dot C(t) = \frac{2\mu^2}{\hbar^2\omega}e^{-{\gamma_0 t}/{\hbar}}\sinh\omega t, \label{eq:Ctdot}
\end{eqnarray}
with $\omega=\sqrt{\gamma_0^{2}-4\mu^{2}}/\hbar$. For example, $\dot{C} (2.34) \approx 0.0096$ as a maximum value for $\mu=0.1$, $\gamma_0=1$, $\hbar=1$.

In the parameter-tunable quantum circuit shown in Fig.~\ref{fig:QCircSupp}, we input $\ket{0}$ and $\ket{1}$ with the same probability to construct the fully mixed state at $\hat{\rho}_{eq}$. For example, Table~\ref{tab:CTermTable} shows the analytic expressions for four contributions to the correlation function $C(t)$. The equal contribution of $\ket{0}$ and $\ket{1}$ for $\hat{\rho}_{eq}$ is shown in the first column. Moreover, the addition of $\hat{\mathds{1}}$ and $\hat \sigma_z$ reveals the projector representation $\hat{\theta}_A = (\hat{\mathds{1}} + \hat{\sigma}_z)/2$ at the fixed $\hat{\rho}_{eq}$. Thus,  the third and fourth contributions ($\hat{\rho}_{eq} = \ket{1}\bra{1}$ with $\hat{\theta}_A = \hat{\mathds{1}}$ and $\hat{\theta}_A = \hat{\sigma}_z$) exactly cancel each other out exactly while the remaining two contributions, $\hat{\rho}_{eq} = \ket{0}\bra{0}$ with $\hat{\theta}_A = \hat{\mathds{1}}$ and $\hat{\theta}_A = \hat{\sigma}_z$, yield identical contributions. Therefore, the overall contribution allows us to perform the quantum circuit only with $\hat{\rho}_{eq} = \ket{0}\bra{0}$ to compute $C(t)$.

Moreover, Table \ref{tab:TermTableSupp} gives analytic expressions for all terms contributing to the time derivative of the correlation function $\dot{C} (t)$. Again, we use the decomposition of $\hat{\theta}_A = (\hat{\mathds{1}} + \hat{\sigma}_z)/2$. For this case, we observe cancellation between the jump and anti-commutator contributions in Eq.~\eqref{eq:dCdtSchro} while the Hamiltonian parts $\mathcal{E}_{H1}$ and $\mathcal{E}_{H2}$ give identical contributions. Thus, we can calculate the transition rate as
\begin{equation}
    \dot{C}(t) = \frac{2}{\hbar} \mathcal{E}_{H1}(t).
\end{equation}

In Fig.~\ref{fig:spin-half01}, numerical QuTiP emulations of our method using the circuitry are compared with analytical results $\mu=0.1$, $\gamma_0=1$, $\hbar=1$ in Eqs.~(\ref{eq:Ct001}) and (\ref{eq:Ctdot})~\cite{johansson2012qutip}. Each emulation runs for $N= 3$, 10, 25 time steps, recycling a single ancillary qubit inside the CW-Lindblad time evolution $\mathrm e^{\mathcal L t}$. The discrepancy decreases systematically with finer time discretization since larger $N$ implies shorter time step $\delta=t/N$ in $e^{{\mathcal L} t}$ (see more details in the Appendix). 

\begin{table}
\centering
\resizebox{\linewidth}{!}{%
\begin{tabular}{|c|c|c|c|c|c|}
\hline
$\mathcal{E}$ & $\chi$ & $\hat{N}$ & $\hat{M}$ & $\hat{\theta}_A$ & $\mathcal{E}$ \\
\hline
\multirow{2}{*}{$\mathcal{E}_{H1}$} & \multirow{2}{*}{$-\frac{\pi}{2}$} & \multirow{2}{*}{$\hat{\mathds{1}}$} & \multirow{2}{*}{$\hat{H}=\mu \hat{\sigma}_y$} & $\hat{\mathds{1}}$ & $0$ \\
 &  &  &  & $\hat{\sigma}_z$ & $\frac{\mu^2}{\hbar \, \omega} e^{- \gamma_0 t/\hbar} \sinh(\omega t)$ \\
\hline
\multirow{2}{*}{$\mathcal{E}_{H2}$} & \multirow{2}{*}{$\frac{\pi}{2}$} & \multirow{2}{*}{$\hat{H}=\mu \hat{\sigma}_y$} & \multirow{2}{*}{$\hat{\mathds{1}}$} & $\hat{\mathds{1}}$ & $0$ \\
 &  &  &  & $\hat{\sigma}_z$ & $\frac{\mu^2}{\hbar\, \omega} e^{- \gamma_0 t/\hbar} \sinh(\omega t)$ \\
\hline
\multirow{2}{*}{$\mathcal{E}_{J}$} & \multirow{2}{*}{$0$} & \multirow{2}{*}{$\hat{L}^{\dagger}=\sqrt{\gamma_0} \hat{\sigma}_z$} & \multirow{2}{*}{$\hat{L}=\sqrt{\gamma_0}\hat{\sigma}_z$} & $\hat{\mathds{1}}$ & $\gamma_0/ 2~~$ \\
 &  &  &  & $\hat{\sigma}_z$ & $-\frac{\gamma_0}{2 } e^{- \gamma_0 t/\hbar} f (t) ^{ }$ \\
\hline
\multirow{2}{*}{$\mathcal{E}_{AC1}$} & \multirow{2}{*}{$0$} & \multirow{2}{*}{$\hat{\mathds{1}}$} & \multirow{2}{*}{$\hat{L}^{\dagger}\hat{L}=\gamma_0 \hat{\mathds{1}}$} & $\hat{\mathds{1}}$ & $\gamma_0/ 2~~$ \\
 &  &  &  & $\hat{\sigma}_z$ & $-\frac{\gamma_0}{2} e^{- \gamma_0 t/\hbar} f (t)$ \\
\hline
\multirow{2}{*}{$\mathcal{E}_{AC2}$} & \multirow{2}{*}{$0$} & \multirow{2}{*}{$\hat{L}^{\dagger}\hat{L}=\gamma_0 \hat{\mathds{1}}$} & \multirow{2}{*}{$\hat{\mathds{1}}$} & $\hat{\mathds{1}}$ & $ \gamma_0/2~~$ \\
 &  &  &  & $\hat{\sigma}_z$ & $-\frac{\gamma_0}{2} e^{- \gamma_0 t/\hbar} f (t)$ \\
\hline
\end{tabular}}
\caption{ Analytic expressions contributing to  $\dot{C}(t)$ in Eq.~\eqref{eq:dCdtSchro} for the spin-1/2 model. The operator \( \hat{\theta}_A \) is implemented in Fig.~\ref{fig:QCirc} by the combination of $\hat{\mathds{1}}$ and $\hat{\sigma}_z$ $\left( f (t) =  \cosh\left(\omega t \right) + \frac{\gamma_0 }{\hbar \omega}\sinh\left(\omega t \right) \right)$.}
\label{tab:TermTableSupp}
\end{table}

\subsection{Quantum circuits on \textit{ibm\_brisbane} }
\label{M-5}
\begin{figure*}[t]
  \centering
  \includegraphics[width=0.8\textwidth]{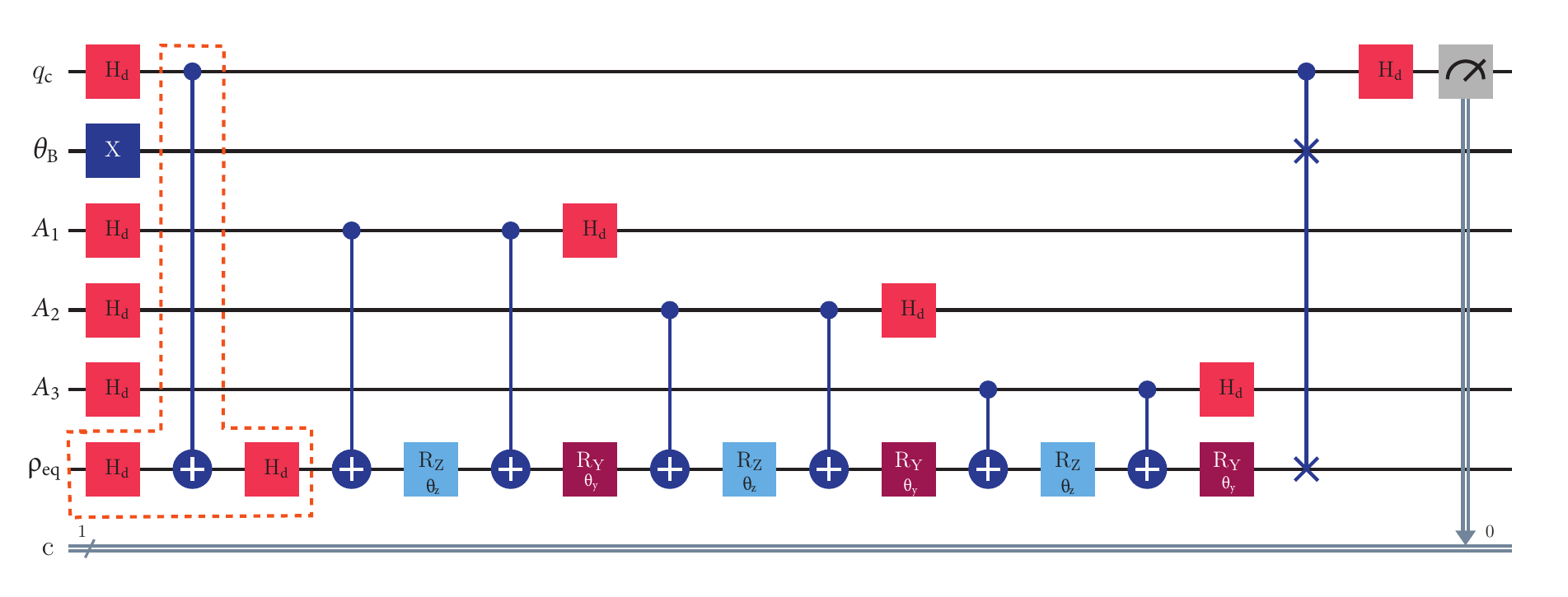}
  \caption{Six-qubit IBMQ circuit for $\mathcal{E}_{C}(t)$ in $C(t)$.  Ancillary qubits $A_1$–$A_3$ implement the
  CW-Lindblad time-evolution gate.  All qubits start in $\ket{0}$ and we choose $\rho_{\rm eq} = \ket{0}\bra{0}$ based on the combination in Table \ref{tab:CTermTable}.
  The gates inside the dashed red box correspond to
  $\hat{\theta}_A=\hat{\sigma}_z$ and are omitted for
  $\hat{\theta}_A=\hat{\mathds{1}}$ due to $\ket{0}\bra{0} = (\hat{\mathds{1}} + \hat{\sigma}_z)/2$ in Table \ref{tab:CTermTable}. 
  Single-qubit rotations in the $\hat{\rho}_{\mathrm{eq}}$ channel are parametrised as
  $\hat R_z(\theta_z)$ in blue boxes and $\hat R_y(\theta_y)$ in purple boxes.}
  \label{fig:CZ1}
\end{figure*}

\begin{figure*}[t]
  \centering
  \includegraphics[width=1\textwidth]{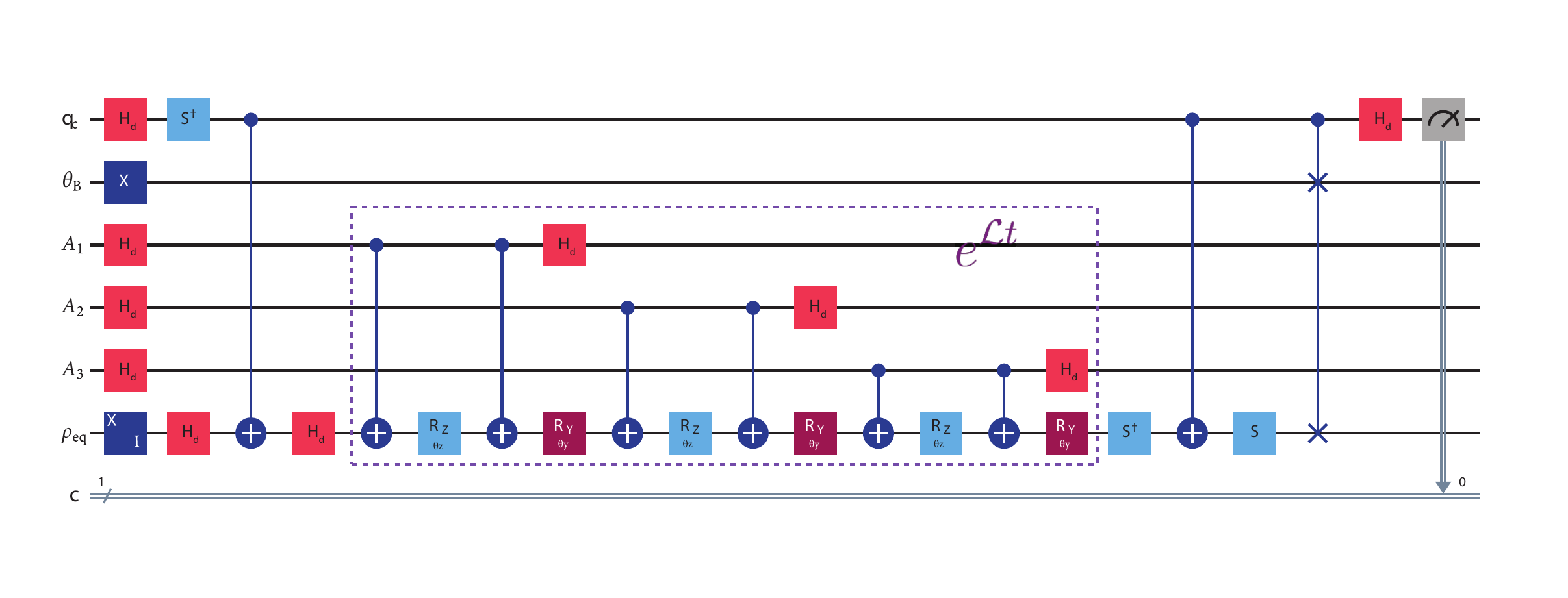}
  \caption{Six-qubit IBMQ circuit with Qiskit used to evaluate the single
  expectation value
  $\mathcal{E}_{\tilde{H}1}(t)$ that determines $\dot{C}(t)$. The dashed box contains the CW-Lindblad evolution
  $e^{\mathcal{L}t}$ realised with the same three ancillas $A_1$–$A_3$.
  The bottom qubit $\hat{\rho}_{\rm eq}$ is prepared in a maximally mixed state by applying
  either $\hat{\sigma}_x$ or $\hat{\mathds{1}}$ with equal
  probability.}
  \label{fig:CH1}
\end{figure*}

For hardware validation, we now execute the quantum simulation algorithm on IBMQ using six qubits: three main qubits for control, $\hat\theta_B = \ket{1} \bra{1}$, and system initialized to $\hat{\rho}_{\rm eq}$, and three ancillas for realizing $N=3$ repeated-interaction steps for $e^{{\mathcal L} t}$ (blue dashed box in Fig.~\ref{fig:CH1}), and the actual IBMQ circuits are shown in Figs.~\ref{fig:CZ1} and \ref{fig:CH1}. For example, for the input of $\rho_{\rm eq}$, we apply the operator either $\mathds{1}$ or $\hat{\sigma}_x$ on $\ket{0}$ with equal probability. For a minimal implementation, the controlled-$\hat{\theta}_A$ gate is decomposed using $\hat{\theta}_A=\ket{0}\bra{0}=(\mathds{1}+\hat{\sigma}_z)/2$, reducing the primitive to a controlled-$\hat\sigma_z$ gate plus an identity branch shown at the red dashed box in Fig.~\ref{fig:CZ1}. 

Figs.~\ref{fig:CZ1} and~\ref{fig:CH1} present the two quantum circuits run on
the \textit{ibm\_brisbane} superconducting processor to obtain the
correlation function $C(t)$ and its time derivative
$\dot{C}(t)$, respectively, from measurements of the control qubit
$q_c$.  Both circuits consume the three ancillary qubits
($A_1$–$A_3$) to realise CW-Lindblad evolution, and this time evolution block $e^{{\mathcal L} t}$ is outlined by a dashed line in Fig.~\ref{fig:CH1}. The single-qubit rotation angles in $\hat R_z(\theta_z)$ (blue box) and $\hat R_y(\theta_y)$ (purple box) are fixed by the Lindbladian parameters according to
\begin{equation}
  \theta_z(t)=2\sqrt{\frac{\gamma_0 \, t}{N}}\, ,
  \qquad
  \theta_y(t)=2\frac{\mu t}{N}\, ,
\end{equation}
where $N$ is the number of ancilla iterations (three qubits in our
experimental circuit). The numerical angle values are listed for fixed $t$ in
Table~\ref{tab:theta_values}.
\begin{table}[h]
\centering
\begin{tabular}{|c|c|c|c|c|c|c|}
    \hline
    $t $ & 0.0 & 0.2 & 0.4 & 0.6 & 0.8 & 1.0 \\
    \hline    \hline
    $\theta_z$ & 0 & $\frac{2}{\sqrt{15}}$ & $\frac{2\sqrt{2}}{\sqrt{15}}$ & $\frac{2\sqrt{3}}{\sqrt{15}}$ & $\frac{4}{\sqrt{15}}$ & $\frac{2\sqrt{5}}{\sqrt{15}}$ \\
    \hline
    $\theta_y$ & 0 & $\frac{1}{75}$ & $\frac{2}{75}$ & $\frac{3}{75}$ & $\frac{4}{75}$ & $\frac{5}{75}$ \\
    \hline
\end{tabular}
\caption{The rotation gate parameters $\theta_z$ and $\theta_y$ are used in Figs.~\ref{fig:CZ1} and \ref{fig:CH1} for the various evolution times. The parameters are given with $\gamma_0=1$ and $\mu=0.1$ for our experimental demonstration.}
\label{tab:theta_values}
\end{table}

For each experimental run, we measure the control qubit in the
$\hat{\sigma}_z$ basis at
$t = 0,\,0.2,\,0.4,\,0.6,\,0.8$, and $1$. This yields ten batches of
$20\,000$ shots for $C(t)$ and five batches of $20\,000$ shots for
$\dot{C}(t)$ as shown in the Appendix. The mean of each batch is reported together with its
standard error, and the experimental results are plotted in
Fig.~\ref{fig:IBM_Q_Experiment}.

\subsection{Experimental results for $C(t)$ and $\dot C(t)$}

\begin{figure*}[t]
\centering
\begin{minipage}[b]{0.43\linewidth}
  \includegraphics[width=\textwidth,trim=0cm 0cm 1cm 0cm]{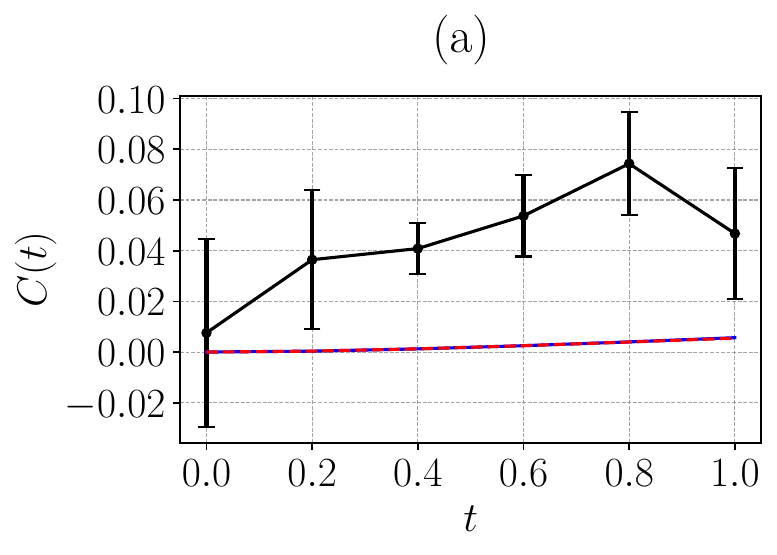}
\end{minipage}\hfill
\begin{minipage}[b]{0.45\linewidth}
  \includegraphics[width=\textwidth,trim=1cm 0cm 0cm 0cm]{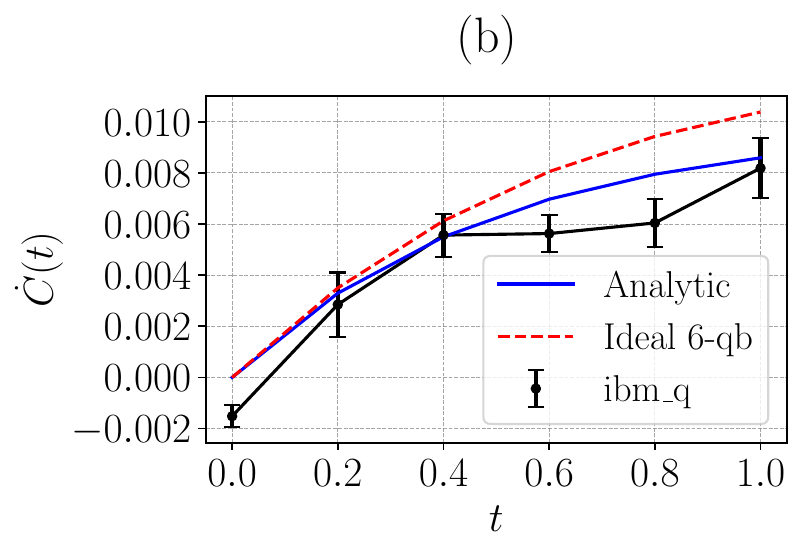}
\end{minipage}
\caption{IBM quantum hardware (six-qubit circuit realization). (a) $C(t)$ from ten batches of $20{\,}000$ shots and (b) $\dot C(t)$ from five batches of $20{,}000$ shots. Both are compared with analytics (blue, solid) and ideal six-qubit simulation (red, dashed) at $t=\{0,0.2,0.4,0.6,0.8,1.0\}$. }
\label{fig:IBM_Q_Experiment}
\end{figure*}

We first estimate $C(t)$ from ten independent batches of $20{\,}000$ shots at $t=\{0,0.2,0.4,0.6,0.8,1.0\}$ in Fig.~\ref{fig:IBM_Q_Experiment} (a). The estimated $C(t)$ increases with $t$ and lies above ideal references. 
Note that the normalization ${\mathcal E}_D=\tr(\hat\rho_{\rm eq} \hat\theta_A)=1/2$ is set for this case. 
Fig.~\ref{fig:IBM_Q_Experiment}(b) shows the data of $\dot C(t)$ from five batches of $20{\,}000$ shots and compares them with Eqs.~(\ref{eq:Ct001}) and (\ref{eq:Ctdot}) in addition to an ideal six-qubit simulation in Qiskit. Apart from a small offset at $t=0$, which is due to the pure gate errors without any time evolution defects, the experimental curve tracks the predicted shape closely. Applying a uniform vertical shift yields excellent agreement with only minor residuals (no error-mitigation applied). Interestingly, $\dot C(t)$ exhibits smaller standard errors than $C(t)$ despite using half as many shots, consistent with
\begin{eqnarray}
\dot C(t)={2 \over \hbar} \mathcal E_{H1}={2 \mu \over \hbar} \mathcal E_{\tilde H1},\qquad \tilde H=\hat\sigma_y,
\label{eq:HtermS1}
\end{eqnarray}
which might suppress variance by the small coupling factor $\mu^2=0.01$ ($\tilde H$: rescaled Hamiltonian without $\mu$). Note that the simplest expectation value arises in Eq.~(\ref{eq:HtermS1}) due to the compensation of the jump and non-Hermitian terms each other in Eq.~(\ref{eq:dCdtSchro}). Numerical shot-noise tests in Qiskit corroborate that both $C(t)$ and $\dot C(t)$ converge toward ideal curves as the shot count increases. The experimental raw data are provided in the Appendix.

\begin{figure*}[t]
\centering
\begin{minipage}[b]{0.44\linewidth}
  \includegraphics[width=\textwidth,trim=0cm 0cm 1cm 0cm]{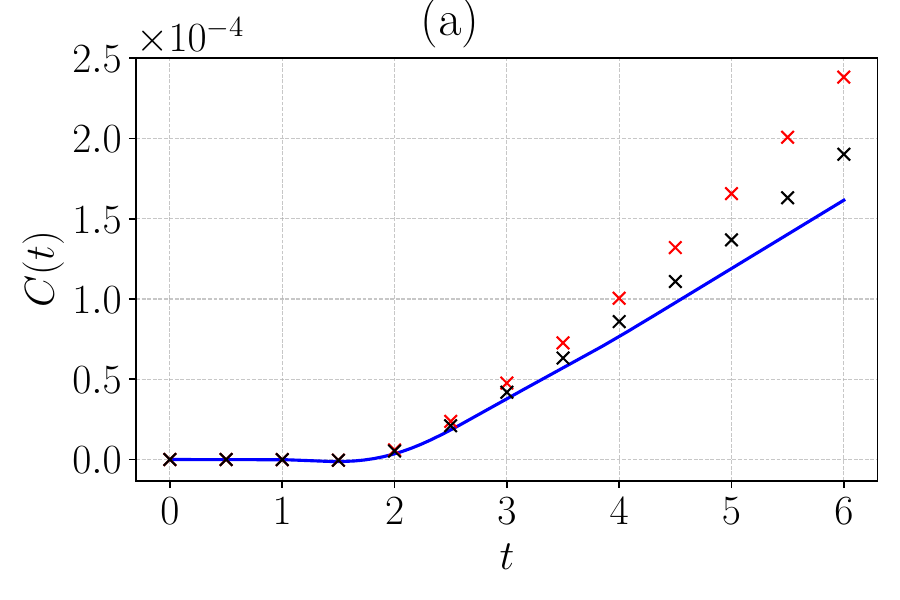}
\end{minipage}\hfill
\begin{minipage}[b]{0.44\linewidth}
  \includegraphics[width=\textwidth,trim=1cm 0cm 0cm 0cm]{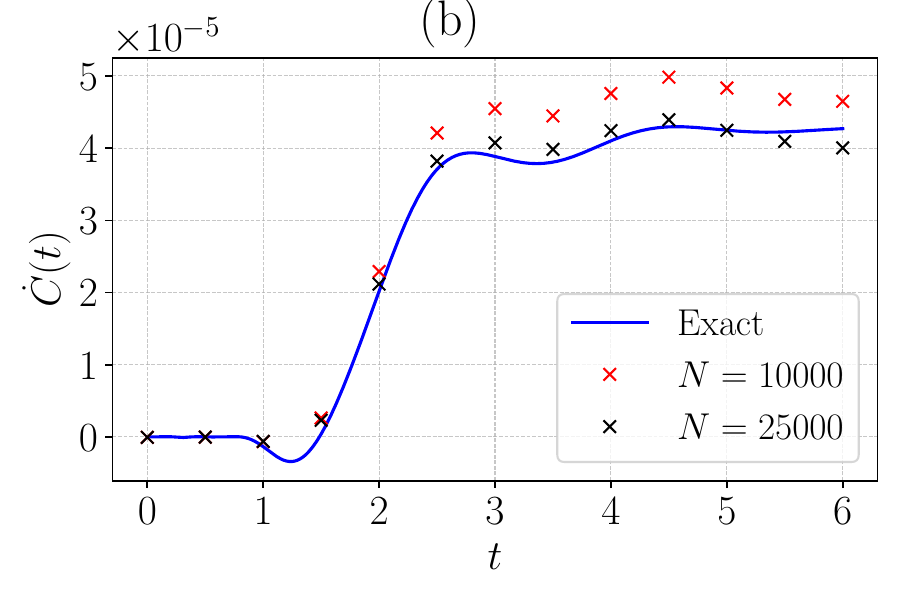}
\end{minipage}
\caption{Caldeira–Leggett double well: (a) $C(t)$ and (b) $\dot C(t)$ for $m=1$, $k_B T=0.0162$, $\epsilon=0.001$, $\hbar=0.01$. Blue solid: direct numerics; markers: QuTiP emulation of Fig.~\ref{fig:QCirc} at $N=10^4$ and $2.5\times 10^4$ steps.}
\label{fig:CaldeiraRate-total}
\end{figure*}

\section{A realistic benchmark: \\ Caldeira–Leggett double-well model}
We next study quantum Brownian motion in a 1D double well known as the Caldeira–Leggett model \cite{CLmodel83}, where activated transport and slow relaxation are crucial and modern numerical schemes can be costly with its system size. In the high-temperature approximation (e.g., $k_B T$ well above the ground-state scale), the Brownian dynamics are captured by a positivity-preserving Lindblad equation~\cite{petruccione} with
\begin{eqnarray}
\hat{H} &=& \frac{\hat{P}^{2}}{2m}+\hat{V}(\hat{X})+\frac{\gamma}{2}\left( \hat{X}\hat{P} + \hat{P}\hat{X} \right), \\
\hat{L} &=& \sqrt{\gamma}\left(\lambda_T^{-1}\hat{X} + i\lambda_T \hat{P}\right),
\label{eq:CL_HL}
\end{eqnarray}
where $\lambda_T=\sqrt{\hbar/(4mk_B T)}$ ($k_B$: Boltzmann constant). 
The Hamiltonian and Lindblad operators for the Caldeira–Leggett model are explained in the Appendix based on Ref.~\cite{Breuer2002}.

The non-unitary observables (e.g., $\hat X$ and $\hat P$) are again implemented via the small-angle approximation with $\epsilon$ in Eq.~\eqref{eq:Ohat_approx}, with accuracy quantified in the Appendix. We discretize the coordinate $x\in[0,1-2^{-n}]$ onto $n$ grid qubits ($\delta x=2^{-n}$) and encode $\hat\rho_{\rm eq}$ and $\hat\theta_{A/B}$ on the register. 
To initialize $\hat\rho_{\rm eq}$, we can in general consider the propagation of an arbitrary state under $\mathcal L$ until convergence to ${\mathcal L}(\hat\rho_{\rm eq})=0$.
When $\hat\rho_{\rm eq}$ is Gibbsian like in this Caldeira–Leggett model, dedicated Gibbs-state sampler schemes can offer improved asymptotic scaling at the cost of additional preparation~\cite{yung2012quantum,motta2020determining,rall2023thermal}. The projectors are prepared as normalized mixed states,
\begin{eqnarray}
\hspace{-0.5cm} \hat{\rho}_A = {\cal N}_A \, \hat\theta_{A} = {\cal N}_A\sum_{0.125 \le x_k \le 0.25} \ket{x_k}\bra{x_k}, \\
\hspace{-0.5cm} \hat{\rho}_B = {\cal N}_B \, \hat\theta_{B} = {\cal N}_B\sum_{0.75 \le x_k \le 0.875} \ket{x_k}\bra{x_k},
\label{eq:thetaAB}
\end{eqnarray}
where $x_k$ is the $k$th grid point and ${\cal N}_{A/B}$ are pre-known normalization factors with the projector sizes. 
The double-well potential is given in the grid basis,
\begin{eqnarray}
\hat{V}(\hat{X})=20\sum_k (x_k- {1 \over 5})^2(x_k- {4 \over 5})^2\ket{x_k}\bra{x_k},  
\label{eq:doublewell}
\end{eqnarray}
yielding a barrier $V_B=0.162$ in the double well. Note that we set $k_B T=0.1\,V_B$ to probe activated dynamics without fully suppressing tunneling. 

In Fig.~\ref{fig:CaldeiraRate-total}, 11-qubit QuTip emulations use $n=5$ grid qubits (32 grid points) duplicated for $\hat\theta_B$ and $\hat\rho_{\rm eq}$ plus a control qubit and the CW-Lindblad evolution block uses $N=10^4$ and $2.5\times 10^4$ time steps for long-time evolution ($0\le t \le6$). 
Our method for $\dot C(t)$ in Fig.~\ref{fig:CaldeiraRate-total}(b) clearly reproduces both two unique quantum features such as short-time intrawell relaxation ($1\!\lesssim\!t\!\lesssim\!2$) and the plateau near $t\!\approx\!3$ while it is too difficult to determine the linear-response regime from the feature of $C(t)$ in Fig.~\ref{fig:CaldeiraRate-total}(a).
Increasing $N$ again reduces time discretization error during $e^{{\mathcal L} t}$. 
Crucially, unlike conventional direct long-time fits or full Liouville-space propagation, our workflow can guarantee to estimate $\dot C(t)$ from a constant number of single-time expectations.

\section{Discussion}
We introduced and demonstrated a hardware-ready framework for estimating transition rates in open quantum systems by reducing the time derivative of an equilibrium correlation function to a finite sum of single-time expectation values. 
This derivative-to-expectation reduction removes two bottlenecks: explicit Liouville-space propagation and long-time decay fitting. 
All required quantities are obtained from a constant number of independently executable circuit instances, compatible with shallow-depth NISQ devices. 
Our scheme was validated with two critical examples. First, the six-qubit IBMQ realization (spin-$1/2$ dephasing model) reproduced the analytical rate profile, establishing experimental viability of the estimator on contemporary quantum hardware. Second, the 11-qubit Caldeira–Leggett benchmark evidently captured both intrawell relaxation and the late-time plateau. 

We would like to emphasize that a key advantage is resource scaling for real-world problems.
For example, a 3D Caldeira–Leggett system discretized on a $1024^3$ grid needs $2^{30}$ points and a dense classical density operator requires $\sim\,2^{60}$ complex entries at a fixed time, already over TiB-scale memory, while manipulating the superoperator is prohibitive even with sparsity. 
By contrast, our circuitry needs only $2n+1$ qubits to obtain $C(t)$ or $\dot C(t)$, where $n=\log_2(1024^3)=30$ qubits encode the grid, and 61 qubits (two registers for $\hat\theta_B$ and $\hat\rho_{\rm eq}$ plus one control qubit) are required in total with $\mathcal O(1)$ recyclable ancillas for the CW-Lindblad time evolution.

Furthermore, three developments will potentially strengthen applicability: (i) more faithful processors (e.g., mid-scale qubits, higher gate fidelities, error-mitigation); (ii) algorithmic upgrades for a new modular block $e^{\mathcal L t}$ (e.g., sparse product-formula \cite{childs2017efficient} or amplitude-amplified Lindbladian simulation  \cite{li2023simulating}) to lower depth at fixed accuracy; and (iii) broader dynamical regimes via embeddings or auxiliary channels for non-Markovian settings. These could shift the model complexity into hardware-native primitives in the near future.  \\
\\
{\bf Acknowledgments:} 
We acknowledge support from the Institute for Information \& Communications Technology Promotion (IITP) grant funded by the Korea government (MSIP) (No. 2019-000003). 
J.\,B.\ and K.\,B.\ are supported by the Ministry of Trade, Industry, and Energy (MOTIE), Korea, under the project “Industrial Technology Infrastructure Program” (RS-2024-00466693) and by the National Research Foundation of Korea (NRF-2023M3K5A1094813 and RS-2023-00281456). 
This work is also supported by Grant No.~K25L5M2C2 at the Korea Institute of Science and Technology Information (KISTI). 
We acknowledge utilization of IBM quantum services. 
The views expressed are those of the authors and do not reflect the official policy or position of IBM or the IBM Quantum team. 

{ \bf Data Availability:}
The data supporting this study are presented in the article, and are available from the corresponding authors upon reasonable request. 
\\
{\bf Code Availability:} Qiskit and Qutip codes are available from the corresponding authors upon reasonable request. \\
{\bf Author Contributions:}  
J. J. and R. C. initiated the research and conducted the research. All authors prepared and revised the manuscript and contributed to experiments and discussions.\\
\\
{\bf Competing interests:} The authors declare no competing financial interests.\\

\appendix
\section{Appendix}

\subsection{Transition rate theory}
We first review the classical transition rate theory based on \cite{chandlerbook}, before introducing an analogous quantum formulation in Lindblad dynamics. To define a transition between the ``states'', we first identify metastable subspaces \(A\) and \(B\) within the phase space of the system. We use characteristic functions to define the state functions for regions \(A\) and \(B\) as follows:
\begin{eqnarray}
&&    \theta_A(x) = \begin{cases}
    1, & \text{if} ~~  x \in A,\\
    0, & \text{if} ~~  x \notin A,
  \end{cases} ~~ \text{and} ~~ \nonumber \\ 
&&  \theta_B(x) = \begin{cases}
    1, & \text{if} ~~  x \in B,\\
    0, & \text{if} ~~  x \notin B.
  \end{cases}
\end{eqnarray}
If we consider the case of many non-interacting particles at equilibrium, the populations of states \( A \) and \( B \) fluctuate due to transitions between these states. Consequently, the transition dynamics can be described by the temporal correlation of the states populations from point $x_0$ to point $x_t$ as follows:
\begin{equation}
    C(t) = \frac{\ev{\theta_A(x_0, 0)  \theta_B(x_t,t)}_{eq}}{\ev{\theta_A(x_0,0)}_{eq}}
\end{equation}
In this expression, \( \ev{\cdot}_{eq} \) represents the average over the equilibrium distribution of the initial state. 

The correlation function \( C(t) \) is defined as the conditional probability of observing the system in state \( B \) at time \( t \) assuming that it started from state \( A \) at $t=0$ as shown in Fig.~1 in the main text. As described by linear response theory \cite{chandlerbook,onsager1931reciprocal,berezhkovskii2023population}, the rate of equilibrium fluctuations from \( A \) to \( B \) is equal to the rate of relaxation as the system recovers from a non-equilibrium condition in which only state $A$ is initially occupied. For brief intervals, \( C(t) \) reflects microscopic movements within state A and the transition state area, linked on the molecular time scale \( t_{mol} \) \cite{chandlerbook}, which represents the time required to traverse the barrier separating the stable regions and settle into one of the states. However, on timescales longer than \( t_{mol} \), a two-state kinetic model provides a good description of the transition dynamics. In this model, transitions are rare compared to the time spent in metastable states. The dynamical model can be represented as follows:
\begin{align} 
    \dv{}{t}\ev*{{\theta}_A (t)}_{ne} &= -k_{AB} \ev{{\theta}_A (t) }_{ne}  + k_{BA} ~ \ev{{\theta}_B (t)}_{ne}, \label{eq:Akin} \\
    \dv{}{t}\ev*{{\theta}_B (t)}_{ne}  &= k_{AB} \ev{{\theta}_A (t)}_{ne} - k_{BA} \ev{{\theta}_B (t)}_{ne}. \label{eq:Bkin}
\end{align}
The expectation values are taken with respect to time-dependent non-equilibrium phase-space probability densities, the solutions are
\begin{align}
    \ev{{\theta}_A (t)}_{ne} &= \frac{k_{BA} + k_{AB}  e^{-(k_{AB} + k_{BA})t}}{k_{AB} + k_{BA}}, \\
    \ev{{\theta}_B (t)}_{ne} &= \frac{k_{AB} \left(1 - e^{-(k_{AB} + k_{BA})t}\right)}{k_{AB} + k_{BA}}, \label{eq:CtSol}
\end{align}
which satisfy the kinetic equations with the initial conditions at $t=0$
\begin{equation} \label{eq:InitialConditions}
    \ev{{\theta}_A (0)}_{ne} = 1 \quad \text{and} \quad \ev{{\theta}_B (0)}_{ne} = 0.
\end{equation}

After an initial transient time $t_{mol}$, the population dynamics of the double well system are well described by a two-state kinetic model. 
The initial conditions are equivalent to the definition of \( C(t) \) as the conditional probability of observing the system in state \( B \) at time \( t \) such that
\begin{equation}
    C(t) = \ev{{\theta}_B (t)}_{ne}.
\end{equation}
Expanding Eq.~\eqref{eq:CtSol} for short times (first order in $t$) in the linear response domain, the correlation function becomes $C(t) \approx k_{AB} t$. Therefore, if we start the system in the state \( A \), the gradient of \( C(t) \) gives the rate constant \( \dot{C} (t) \approx k_{AB} \).

The classical rate theory extends to quantum systems by replacing the classical characteristic functions with analogous projection operators as shown in Eq.~(\ref{eq:lindblad}). The quantum correlation function derivative is thus given by
\begin{equation} \label{eq:QDcorrel2}
    \dot {C}_{\mathbb{C}}(t) \equiv \frac{\langle \hat{\theta}_A(0) \dot{\hat{\theta}}_B(t) \rangle_{eq} }{\langle \hat{\theta}_A(0) \rangle_{eq}}.
\end{equation}
and is complex-valued. The real part of Eq.~\eqref{eq:QDcorrel2} is identified as the rate, while the imaginary component contains phase information \cite{craig2004quantum}. Since the projectors are Hermitian, the rate component of the correlation function is given by the anti-commutator expectation value
\begin{equation}
    \dot { C}(t) \equiv \frac{\langle \{\hat{\theta}_A(0), \dot{\hat{\theta}}_B(t) \}\rangle_{eq}}{2\langle \hat{\theta}_A(0) \rangle_{eq}}.
\end{equation}
More explicitly, the above expression may be written with $\hat{\theta}_{A/B} (0) \equiv \hat{\theta}_{A/B}$ and $\dot{\hat{\theta}}_B(t) = \mathcal{L}^{\dagger}\left( \hat{\theta}_B(t) \right) $ in the form 
\begin{equation} \label{eq:rate2}
         \dot{C}(t) = \frac{\Tr( \hat{\rho}_{eq} \left\{ \hat{\theta}_A , \mathcal{L}^{\dagger}\left(e^{ \mathcal{L}^{\dagger}t} (\hat{\theta}_B)\right) \right\} )}{2 \Tr(\hat{\rho}_{eq} \hat{\theta}_A)}.
\end{equation}
where \(\hat{\rho}_{eq}\) is the equilibrium state of the Lindblad dynamics. Then, the time evolution of the observables is governed by the Heisenberg picture Lindblad evolution equation:
\begin{eqnarray} 
 \hspace{-0.8cm} && \frac{d}{dt} {\hat{\theta}}(t)   = \mathcal{L}^{\dagger} \left( \hat{\theta}(t) \right) = \mathcal{L}^{\dagger} \left(e^{ \mathcal{L}^{\dagger}t} (\hat{\theta}_B) \right) \nonumber \\
 \hspace{-0.8cm}  && = \frac{i}{\hbar}[\hat{H},\hat{\theta}(t)]  + \frac{1}{\hbar} \sum_{k}\left(\hat{L}_k^{\dagger} \hat{\theta}(t) \hat{L}_{k } - \frac{1}{2} \{ \hat{\theta}(t) , \hat{L}_k^{\dagger} \hat{L}_k \}\right). \nonumber \\
 \hspace{-0.8cm}  && \label{eq:HeisenLind2}
\end{eqnarray}
By expanding the commutators and rearranging the terms in Eq.~\eqref{eq:rate2}, we can reformulate the transition rate in terms of the Schr\"{o}dinger picture dynamics as follows
\begin{equation} \label{eq:Srate}
         \dot{C}(t) = \frac{\Tr(\hat{\theta}_B \mathcal{L}\left( e^{ \mathcal{L}t} \left( \left\{\hat{\rho}_{eq} , \hat{\theta}_A \right\}\right)\right))}{2 \Tr(\hat{\rho}_{eq} \hat{\theta}_A)}.
\end{equation}
with the standard form of the Lindblad equation in Eq.~(\ref{eq:lindblad}).

\begin{widetext}
\subsection{Mixed states in the correlation derivative circuit} 
We here provide the mathematical details for estimating the expectation value corresponding to a \(\hat{\sigma}_z\) measurement on the control qubit in the correlation derivative circuit in Fig.~\ref{fig:QCirc}. We denote the total density operator at stage $n$ by $\hat{\Phi}_n$. Our initial circuit input is the state
\begin{equation}
    \hat{\Phi}_0 = \ket{+}\bra{+} \otimes \hat{\theta}_B \otimes \hat{\rho}_{eq},
\end{equation}
with the control qubit in the $\ket{+}$ state. We prepare the mixed states (e.g., statstically or using a mixed-state preparation algorithm) described by the projection operator $\hat{\theta}_B$ together with $\hat{\rho}_{eq}$.
In the next stage, we apply a phase-shifted \(z\)-axis rotation operator to the control qubit, such as $\hat{R}(\chi)  = e^{-i \chi (\hat{\sigma}_z -1) /2 }$, and the state \(\hat{\Phi}_1\) is given by
\begin{eqnarray*} \label{eq:nSwapS}
    \hat{\Phi}_1 = \frac{1}{2} \left( \ket{0}\bra{0} + e^{-i \chi} \ket{0}\bra{1} + e^{i \chi} \ket{1}\bra{0} + \ket{1}\bra{1} \right) \otimes \hat{\theta}_B \otimes\hat{\rho}_{eq}.
\end{eqnarray*}
After the controlled-$\hat{\theta}_A$ gate as in Fig.~\ref{fig:QCirc}, $\hat \Phi_2$ is given by
\begin{eqnarray*}
\hat \Phi_2 &=& \frac{1}{2}\bigl(
   \ket{0}\bra{0}\otimes \hat{\theta}_B \otimes \hat{\rho}_{eq}
   + e^{- i \chi} \ket{0}\bra{1}\otimes \hat{\theta}_B \otimes \hat{\rho}_{eq} \hat{\theta}_A \nonumber \\
   && ~~ + e^{i \chi} \ket{1}\bra{0}\otimes \hat{\theta}_B \otimes \hat{\theta}_A \hat{\rho}_{eq}
   + \ket{1}\bra{1}\otimes \hat{\theta}_B \otimes \hat{\theta}_A \hat{\rho}_{eq} \hat{\theta}_A
\bigr).
\end{eqnarray*}
Then, we apply the controlled-$\hat{N}$ gate and the Lindblad time evolution gate $e^{\mathcal{L}t}$ \cite{cleve2016efficient} and the state results in
\begin{eqnarray*}
 \hspace{-2.5cm}    \hat \Phi_3 &=& \frac{1}{2}\bigg(\ket{0}\bra{0}\otimes\hat{\theta}_B \otimes \hat{\rho}_{eq} + e^{- i \chi}\ket{0}\bra{1}\otimes \hat{\theta}_B \hat{N}^{\dagger}\otimes e^{\mathcal{L}t}\left( \hat{\rho}_{eq} \hat{\theta}_A\right)
    \nonumber \\
  \hspace{-2.5cm}   && + e^{i \chi} \ket{1}\bra{0}\otimes \hat{N} \hat{\theta}_B  \otimes e^{\mathcal{L}t}\left(\hat{\theta}_A\hat{\rho}_{eq}\right) + \ket{1}\bra{1}\otimes \hat{N} \hat{\theta}_B \hat{N}^{\dagger} \otimes  \hat 
 e^{\mathcal{L}t}\left(\hat{\theta}_A \hat{\rho}_{eq}  \hat{\theta}_A\right)\bigg).
\end{eqnarray*}
Subsequently, applying the controlled-$\hat{M}$ gate yields
\begin{eqnarray*}
\hat \Phi_4 &=& \frac{1}{2}\bigl(
  \ket{0}\bra{0}\otimes \hat{\theta}_B \otimes \hat{\rho}_{eq} + e^{-i\chi} \ket{0}\bra{1}\otimes \hat{\theta}_B \hat{N}^{\dagger}\otimes
    e^{\mathcal{L}t} \bigl(\hat{\rho}_{eq} \hat{\theta}_A\bigr) \hat{M}^{\dagger} \nonumber \\
&& + e^{i\chi} \ket{1}\bra{0}\otimes \hat{N} \hat{\theta}_B \otimes
    \hat{M} e^{\mathcal{L}t} \bigl(\hat{\theta}_A \hat{\rho}_{eq}\bigr)
  + \ket{1}\bra{1}\otimes \hat{N} \hat{\theta}_B \hat{N}^{\dagger}\otimes
    \hat{M} e^{\mathcal{L}t} \bigl(\hat{\theta}_A \hat{\rho}_{eq} \hat{\theta}_A\bigr) \hat{M}^{\dagger}
\bigr).
\end{eqnarray*}
After the block C-SWAP gates \cite{joo2023commutation}, the density operator becomes 
\begin{eqnarray*}
\hat \Phi_5 &=& \frac{1}{2}\bigl(
  \ket{0}\bra{0}\otimes \hat{\theta}_B \otimes \hat{\rho}_{eq} + e^{-i\chi} \ket{0}\bra{1}\otimes \hat{\theta}_B \hat{N}^{\dagger}
    \xleftrightarrow[\text{b}]{\otimes}
    e^{\mathcal{L}t} \bigl(\hat{\rho}_{eq} \hat{\theta}_A\bigr) \hat{M}^{\dagger} \nonumber \\
&&  + e^{i\chi} \ket{1}\bra{0}\otimes \hat{N} \hat{\theta}_B
    \xleftrightarrow[\text{k}]{\otimes}
    \hat{M} e^{\mathcal{L}t} \bigl(\hat{\theta}_A \hat{\rho}_{eq}\bigr)
  + \ket{1}\bra{1}\otimes \hat{M} e^{\mathcal{L}t} \bigl(\hat{\theta}_A \hat{\rho}_{eq} \hat{\theta}_A\bigr) \hat{M}^{\dagger}
    \otimes \hat{N} \hat{\theta}_B \hat{N}^{\dagger} \bigr). \nonumber \\ 
&&
\end{eqnarray*}

We finally have the final-stage state after the Hadamard gate such as
\begin{eqnarray*}
\hat \Phi_6 &=& \frac{1}{2}\bigl(
  \ket{+}\bra{+}\otimes \hat{\theta}_B \otimes \hat{\rho}_{eq} + e^{-i\chi} \ket{+}\bra{-}\otimes \hat{\theta}_B \hat{N}^{\dagger}
    \xleftrightarrow[\text{b}]{\otimes}
    e^{\mathcal{L}t} \bigl(\hat{\rho}_{eq} \hat{\theta}_A\bigr) \hat{M}^{\dagger} \nonumber \\
 &&  + e^{i\chi} \ket{-}\bra{+}\otimes \hat{N} \hat{\theta}_B
    \xleftrightarrow[\text{k}]{\otimes}
    \hat{M} e^{\mathcal{L}t} \bigl(\hat{\theta}_A \hat{\rho}_{eq}\bigr)
  + \ket{-}\bra{-}\otimes
    \hat{M} e^{\mathcal{L}t} \bigl(\hat{\theta}_A \hat{\rho}_{eq} \hat{\theta}_A\bigr) \hat{M}^{\dagger}
    \otimes \hat{N} \hat{\theta}_B \hat{N}^{\dagger} \bigr). \nonumber \\ 
&&
\end{eqnarray*}
From performing the measurements in the control qubit, we have the expectation value 
\begin{eqnarray*}
\Tr(\hat{\sigma}_z^{c} \hat \Phi_6)=\frac{1}{2}\bigg( e^{- i \chi} \Tr(  \hat{\theta}_B \hat{N}^{\dagger} e^{\mathcal{L}t}\left( \hat{\rho}_{eq} \hat{\theta}_A\right) \hat{M}^{\dagger}) + e^{i \chi}  \Tr(\hat{\theta}_B \hat{M} e^{\mathcal{L}t}\left(\hat{\theta}_A\hat{\rho}_{eq}\right)\hat{N})\bigg),
\end{eqnarray*}
where the superscript c in $\hat{\sigma}_z^c$ indicates the $\hat{\sigma}_z$ Pauli operator on the control qubit. 
Based on the single circuit Fig.~\ref{fig:QCirc}, we can compute all the expectation values for Eqs.~(\ref{eq:dCdtSchro-1}) and (\ref{eq:dCdtSchro}). 

\subsection{Cleve-Wang Lindblad time evolution circuit}
\label{sec:CleveLind}

\begin{figure}[b]
\centering 
\begin{quantikz}
 \lstick{$\hat{\rho}_n$}  \qw & \gate{e^{\mathcal{L} \delta}} &\qw \rstick{$\hat{\rho}_{n+1}$} 
\end{quantikz}
$\quad  \equiv  \quad$ 
\begin{quantikz}
~~~~~  & \lstick{$\hat{\rho}_A$} & \gate[2]{e^{- i J\sqrt{\delta}}} & \qw \rstick{trace out} \\
~~~~~  & \lstick{$\hat{\rho}_n$} & \qw & \qw & \gate{e^{- i \hat{H}\delta}} & \qw \rstick{$\hat{\rho}_{n+1}$}
\end{quantikz}
\caption{Schr\"{o}dinger picture time-evolution circuit\label{fig:CLeveProp} \cite{cleve2016efficient}}
\end{figure}

We recap the repeated interaction Lindblad simulation scheme described by Cleve and Wang in \cite{cleve2016efficient}. In their approach, time is discretised into \( N \) segments and an ancillary qudit with dimension \( d + 1 \) is required where \( d \) is the number of Lindblad operators. The qudit state $\hat{\rho}_A$ is initially prepared in
\begin{equation}
    \hat{\rho}_A = \ket{0_{d+1}}\bra{0_{d+1}}.
\end{equation}
Using qubit architectures, if \( d + 1 \) is not a power of two, the qudit can be embedded into a qubit-based register with dimension \( 2^n \geq d + 1 \) for $n$ qubits. The combined system undergoes the joint unitary evolution under the Hamiltonian-like operator
\begin{equation}
    \hat{J} = \begin{pmatrix}
    0 & \hat{L}_1^{\dagger} & \cdots & \hat{L}_d^{\dagger} \\
    \hat{L}_1 & \ddots & & 0 \\
    \vdots & & \ddots & \vdots \\
    \hat{L}_d & 0 & \cdots & 0 
    \end{pmatrix},
    \label{J}
\end{equation}
for the duration of \( \sqrt{\delta} \), where $\delta = {t}/{N}$. After the ancillary qubit is traced out as shown in Fig.~\ref{fig:CLeveProp}, the remaining reduced density matrix is evolved under \( \hat{H} \) for the time step \( \delta \). This process is repeated $N$ times to simulate the approximate Lindblad dynamics with accuracy increasing with $N$. 

For the first stage, in block matrix notation, the total input state is given by
$\ket{0_{d+1}}\bra{0_{d+1}}\otimes \hat{\rho}$. The joint unitary evolution is given by
\begin{eqnarray}
\hspace{-1cm} &&  e^{-i \hat J\sqrt{\delta}}\left(\ket{0_{d+1}}\bra{0_{d+1}}\otimes \hat{\rho}\right) e^{i \hat J\sqrt{\delta}} \nonumber \\
\hspace{-1cm} && ~~ \approx \ket{0_{d+1}}\bra{0_{d+1}}\otimes \hat{\rho}-i\left[\hat{J}, \left(\ket{0_{d+1}}\bra{0_{d+1}}\otimes \hat{\rho}\right)\right]\sqrt{\delta } \nonumber \\ 
\hspace{-1cm} && ~~~~~ +\Bigg(\hat{J}\left(\ket{0_{d+1}}\bra{0_{d+1}}\otimes \hat{\rho}\right)\hat{J}-\frac12\hat{J}^{2}\left(\ket{0_{d+1}}\bra{0_{d+1}}\otimes \hat{\rho}\right) -\frac12\left(\ket{0_{d+1}}\bra{0_{d+1}}\otimes \hat{\rho}\right)\hat{J}^{2}\Bigg)\delta  +O(\delta^{3/2}). \nonumber \\ 
\hspace{-1cm} &&
\end{eqnarray}
The only non-zero entries of the commutator term are off-diagonal in the tensor product representation. The reduced density matrix after tracing out the ancillary channel can be evolved under the Hamiltonian $\hat{H}$, giving 
\begin{eqnarray}
\hspace{-2cm} \hat{\rho} \to \hat{\rho} - i[\hat{H}, \hat{\rho}]\delta + \sum_{k=1}^{d}\left(\hat{L}_k\hat{\rho} \hat{L}_k^{\dagger}  - \frac{1}{2}\hat{L}_k^{\dagger} \hat{L}_k\hat{\rho} - \frac{1}{2}\hat{\rho} \hat{L}_k^{\dagger} \hat{L}_k \right)\delta + O(\delta^{3/2}).
\end{eqnarray}
Taking this output density matrix and feeding it back into the Lindblad evolution circuit for \(N\) iterations, it approximately becomes the time-evolved density matrix \(\hat{\rho}(t)\).

\subsection{Controlled-SWAP gate for mixed states} \label{sec:C_Swap}
In this section we define a controlled-SWAP (C-SWAP) for two density operators on $\hat{A} \otimes \hat{B}$. Suppose we have the following two density operators such as
\begin{equation}
    \hat{A} = \sum_{ij} a_{ij} \ket{i} \bra{j}, \quad \text{and} \quad \hat{B} = \sum_{kl} b_{kl} \ket{k} \bra{l}.
\end{equation}

The control qubit can be expressed in four components such as 
   $\ket{0}\bra{0},  \ket{0}\bra{1},  \ket{1}\bra{0},$ 
   and  $\ket{1} \bra{1}$.
The first and last components, when combined with the C-SWAP circuit, result in $\ket{0}\bra{0} \otimes \hat{A} \otimes \hat{B}$ and  $\ket{1}\bra{1} \otimes \hat{B} \otimes \hat{A}$. If we consider the control input $\ket{1}\bra{0}$, the outcome is given by
\begin{eqnarray} \label{eq:c_swap1}
\hspace{-1cm} \textbf{C}_{\text{SWAP}}\left(\ket{1}\bra{0} \otimes \left(\sum_{ijkl} a_{ij} b_{kl} \ket{i} \bra{j} \otimes \ket{k} \bra{l}\right)\right) = \ket{1}\bra{0} \otimes \left(\sum_{ijkl} a_{ij} b_{kl} \ket{k} \bra{j} \otimes \ket{i} \bra{l}\right). \nonumber \\
\hspace{-1cm} 
\end{eqnarray}
Similarly, for the control input $\ket{0}\bra{1}$, we have
\begin{eqnarray} \label{eq:c_swap2}
\hspace{-1cm} \textbf{C}_{\text{SWAP}}\left(\ket{0}\bra{1} \otimes \left(\sum_{ijkl} a_{ij} b_{kl} \ket{i} \bra{j} \otimes \ket{k} \bra{l}\right)\right) = \ket{0}\bra{1} \otimes \left(\sum_{ijkl} a_{ij} b_{kl} \ket{i} \bra{l} \otimes \ket{k} \bra{j}\right). \nonumber \\
\hspace{-1cm} 
\end{eqnarray}
These states are entangled and cannot be simply expressed in terms of the original operators $\hat{A}$ and $\hat{B}$. However, if we take the partial trace over the $\hat{A}$ and $\hat{B}$ channels, the resulting states are
\begin{equation}
   \Tr(\hat{A} \hat{B}) \ket{1}\bra{0} \quad \text{and} \quad \Tr(\hat{A} \hat{B}) \ket{0}\bra{1}
\end{equation}
respectively for each choice of control qubit. In this way, the C-SWAP operation allows the multiplication of the two channels. To allow us to keep track of swapped states without expanding into a basis, we define a new notation for the state as referenced in Eq.~\eqref{eq:c_swap1} as follows
\begin{equation}
    \sum_{ijkl} a_{ij} b_{kl} \ket{k} \bra{j} \otimes \ket{i} \bra{l} = \hat{A} \xleftrightarrow[\text{k}]{\otimes} \hat{B},
\end{equation}
where the ``k'' signifies that we are swapping the ket parts of the projectors. Similarly, for the state referenced in Eq.~\eqref{eq:c_swap2}, we have
\begin{equation}
    \sum_{ijkl} a_{ij} b_{kl} \ket{i} \bra{l} \otimes \ket{k} \bra{j} = \hat{A} \xleftrightarrow[\text{b}]{\otimes} \hat{B}
\end{equation}
where ``b'' denotes a bra part swap. After the C-SWAP gate, we have the output states 
\begin{align}
    \ket{0}\bra{0}\otimes\hat A\otimes \hat B\otimes \hat C &\to \ket{0}\bra{0}\otimes\hat A\otimes \hat C\otimes \hat B, \\
    \ket{1}\bra{1}\otimes\hat A\otimes \hat B\otimes \hat C &\to \ket{1}\bra{1}\otimes\hat B\otimes \hat A\otimes \hat C, \\
    \ket{0}\bra{1}\otimes\hat A\otimes \hat B\otimes \hat C &\to \ket{0}\bra{1}\otimes\left(\hat A\xleftrightarrow[\text{b1}]{\otimes} \hat B\right)\xleftrightarrow[\text{b2}]{\otimes} \hat C, \\
    \ket{1}\bra{0}\otimes\hat A\otimes \hat B\otimes \hat C &\to \ket{1}\bra{0}\otimes\left(\hat A\xleftrightarrow[\text{k1}]{\otimes} \hat B\right)\xleftrightarrow[\text{k2}]{\otimes} \hat C. 
\end{align}
Taking the trace-out of the state of $\ket{0}\bra{1}$, we have
\begin{align}
    \Tr(\sum_{ijklmn} a_{ij} b_{kl} c_{mn} \ket{i} \bra{l} \otimes \ket{m} \bra{j}\otimes\ket{k} \bra{n})=\Tr(\hat B \hat A \hat C ).
\end{align}
Similarly, the trace-out of the state of $\ket{1}\bra{0}$ gives
\begin{align}
    \Tr(\sum_{ijklmn} a_{ij} b_{kl} c_{mn} \ket{k} \bra{j} \otimes \ket{i} \bra{n}\otimes\ket{m} \bra{l})=\Tr(\hat A \hat B \hat C).
\end{align}

\subsection{Caldeira-Leggett model for quantum Brownian motion in a 1D double well}

Caldeira and Leggett derived the master equation that governs the evolution of the reduced density operators of a quantum particle coupled to a bath of thermal oscillators~\cite{CLmodel83}. This Caldeira-Leggett master equation is expressed with $\hat{H}_{CL} = {\hat{P}^2}/{(2 m)} + V(\hat{X})$
\begin{eqnarray} 
\frac{d}{dt} \hat{\rho}_s &=& -\frac{i}{\hbar}[\hat{H}_{CL}, \hat{\rho}_s] - \frac{i \gamma}{\hbar}[\hat{X}, \{\hat{P}, \hat{\rho}_s\}] - \frac{2 m \gamma k_B T}{\hbar^2} [\hat{X}, [\hat{X}, \hat{\rho}_s]] ,\label{eq:CLMaster}
\end{eqnarray}
where $\hat X$ and $\hat P$ are the position and momentum operators of the quantum particle.

The above CL dynamics does not unfortunately maintain the positivity of the density matrix, especially at reduced temperatures. Addressing this limitation, a modification in \cite{petruccione} introduces a new term such as
\begin{equation}
-\frac{\gamma}{8 m k_B T}[\hat{P},[\hat{P},\hat{\rho}_s]],
\end{equation}
which becomes insignificant at elevated temperatures. This ensures the positivity of the density matrix and allows the dynamics to be recast in the positivity preserving Lindblad form \cite{gorini1976,lindblad1976generators}  
\begin{equation} 
\frac{d}{dt}\hat{\rho}_S =-\frac{i}{\hbar} \left[ \hat{H}_{} , \hat{\rho}_S \right] + \frac{1}{\hbar} \left(\hat{L}_{} \hat{\rho}_S \hat{L}_{}^{\dagger} -\frac{1}{2} \{ \hat{\rho}_S , \hat{L}_{}^{\dagger} \hat{L}_{}  \} \right),
\end{equation}
with $\lambda_T = {\sqrt{\hbar / 4  m k_B T }}$. Its Hamiltonian and Lindblad operators are 
\begin{eqnarray} \label{eq:CLHL2}
\hat{H}_{} &=& \frac{\hat{P}^2}{2m}+ \hat V (\hat X) +\frac{\gamma}{2}(\hat{X}\hat{P}+\hat{P}\hat{X}),\\ 
\hat{L}_{}&=& \sqrt{\gamma  } \left( \lambda_T^{-1} \hat{X}+i \lambda_T \hat{P} \right).
\end{eqnarray}
where $\hat X$ and $\hat P$ are the position and momentum operators.

\subsection{Unitary implementation of position space}

For the Caldeira Leggett discrete position representation and the wave function with \( n \) qubits is given over the coordinate interval \( x \in [0, 1 - 2^{-n}] \) using QuTiP \cite{johansson2012qutip}. The position operator is defined as
\begin{equation}
    \hat{X} = \sum_{k=0}^{2^{n}-1} x_k \ket{x_k}\bra{x_k},
\end{equation}
This position space grid induces a momentum grid that satisfies the discrete uncertainty relation, known as the error-disturbance uncertainty relation \cite{kabernik2021transition}, $ \delta x \delta p = {2 \pi \hbar}/{2^{n}}$ implying momentum spacing \( \delta p = 2 \pi \hbar \). Then, the momentum eigenstates have eigenvalues \( p_j = (-2^{n-1} + j)\delta p \) for \( j = 0, 1, \dots, 2^{n}-1 \) and the momentum operator is defined as
\begin{equation}
    \hat{P} = \sum_{j=0}^{2^{n}-1} p_j \ket{p_j}\bra{p_j} = \sum_{j,k,l=0}^{2^{n}-1} p_j \tilde{U}_{jk} \ket{x_k}\bra{x_l} \tilde{U}^{\dagger}_{lj},
\end{equation}
where the unitary matrix \( \tilde{U} \) is related to the quantum Fourier transform:
\begin{equation}
    \tilde{U}_{jk} = \frac{1}{\sqrt{2^n}} \exp\left(\frac{i}{\hbar} x_j p_k \right).
\end{equation}

We adopt the $n$-qubit grid representation $x\in[0,1-2^{-n}]$ used in QuTiP \cite{johansson2012qutip}.  
The position operator is  
\begin{equation}
\hat X=\sum_{k=0}^{2^{n}-1}x_k\ket{x_k}\bra{x_k},
\label{eq:Xop}
\end{equation}
and the associated momentum spacing $\delta p=2\pi\hbar$ satisfies the discrete uncertainty relation  
\begin{equation}
\delta x \delta p=\frac{2\pi\hbar}{2^{n}},\qquad\delta x=2^{-n},
\label{eq:uncertainty}
\end{equation}
with $p_j=(-2^{n-1}+j)\delta p$ $(j=0,\dots,2^{n}-1)$ the momentum operator reads  
\begin{equation}
\hat P=\sum_{j=0}^{2^{n}-1}p_j\ket{p_j}\bra{p_j}
      =\sum_{j,k,l=0}^{2^{n}-1}p_j U_{jk}\ket{x_k}\bra{x_l}U^{\dagger}_{lj},
\label{eq:Pop}
\end{equation}
where the quantum Fourier transform matrix is  
\begin{equation}
U_{jk}=\frac{1}{\sqrt{2^{n}}}\exp\bigl(i\, x_j \, p_k/\hbar\bigr).
\label{eq:QFT}
\end{equation}

Any Hermitian operator $\hat{\mathcal{O}}\in\{\hat\theta_A,\hat X,\hat P\}$ is approximated by a small-angle unitary
\begin{equation}
\hat{\mathcal{O}}_{\epsilon} \simeq\frac{U(\hat{\mathcal{O}},\epsilon)-U(\hat{\mathcal{O}},-\epsilon)}{2 i\epsilon}\quad\text{with}\quad
U(\hat{\mathcal{O}},\epsilon)=\exp\bigl(i\epsilon\,\hat{\mathcal{O}}\bigr).
\label{eq:finitediff}
\end{equation}
\begin{figure}
  \centering
  \begin{minipage}[b]{0.15\textwidth}
    \centering
    \includegraphics[width=\textwidth,trim=-1cm 0cm 2cm 0cm]{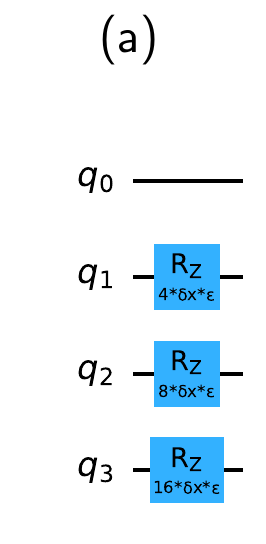}
  \end{minipage}
  \hfill
  \begin{minipage}[b]{0.7\textwidth}
    \centering
    \includegraphics[width=\textwidth,trim=1.5cm 0cm 0cm 0cm]{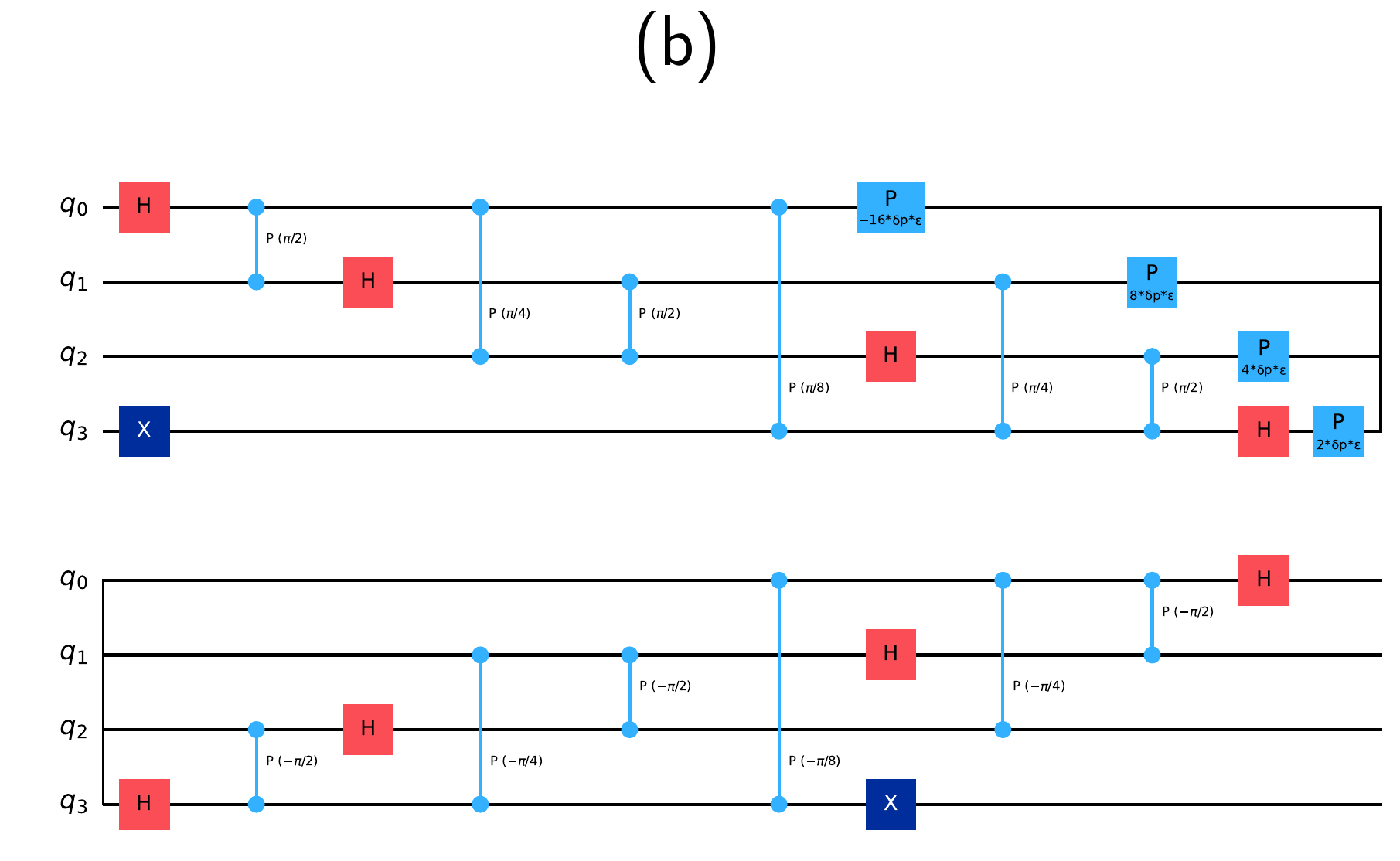}
  \end{minipage}
  \caption{Four-qubit Qiskit implementation of (a) $U(\hat{X},\epsilon)$ and (b) $U(\hat{P},\epsilon)$. Extra NOT gates on the most significant bit in (b), map the domain from $[0,1)$ to $[-\tfrac12,\tfrac12)$ and back. \label{fig:XPUnitary}}
\end{figure}
\begin{figure}
\centering 
\begin{quantikz}
 \lstick{$\ket{\psi}$} \qw & \gate{e^{i \epsilon  \hat \theta_A}} & \qw \rstick{$\ket{\phi}$}
\end{quantikz}
$\quad  \equiv  \quad$
\begin{quantikz}
    \lstick{$q_0$} & & \qw & \qw\\
    \lstick{$q_1$} & & \qw & \qw  \\
    \vdots\\
    \lstick{$q_n$} & \gate{X} & \gate{R_z(\epsilon)} & \qw 
\end{quantikz}
\caption{Unitary used in $\hat \theta_A$ approximation. The NOT and $R_z(\epsilon)$ are applied to the most significant qubit affecting only the smallest $2^{n-1}$ gridpoints \label{fig:AUnitary}}
\end{figure}
\begin{figure}
  \centering
  \includegraphics[width=0.65\textwidth]{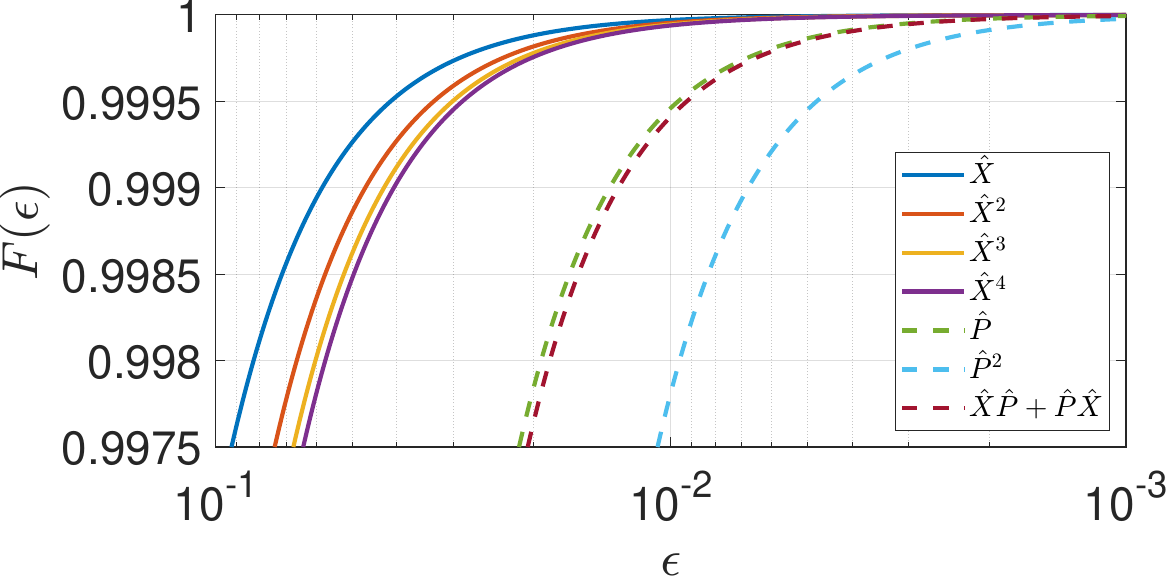}
  \caption{Gate fidelities of the unitary approximations as a function of $\epsilon$ for simulations performed in a truncated five‑qubit
Hilbert space ($d=2^5=32$) representing a uniform grid over the interval $[0,1)$.}
  \label{fig:gateFidelity}
\end{figure}
In Figs.~\ref{fig:XPUnitary} and \ref{fig:AUnitary} we show basic unitary implementations of
\(U(\hat X,\epsilon)\), \(U(\hat P,\epsilon)\) and \(U(\hat \theta_{A,\epsilon}\). We use \(U(\hat X,\epsilon)\) and \(U(\hat P,\epsilon)\) as building blocks to generate approximate unitary decompositions of arbitrary
powers and combinations of \(\hat X\) and \(\hat P\). For small \(\epsilon\),
a sequence of finite differenceidentities provides the required operator
reconstructions given explicitly by
\begin{eqnarray}
  \hat{X} &\approx \frac{U(\hat{X},\epsilon)-U(\hat{X},-\epsilon)}{2 i\epsilon}, \\
  \hat{P} &\approx \frac{U(\hat{P},\epsilon)-U(\hat{P},-\epsilon)}{2 i\epsilon}, \label{eq:X_P_1st}\\
  \hat{X}^2 &\approx -\frac{U(\hat{X},\epsilon)+U(\hat{X},-\epsilon)-2\hat{\mathds{1}}}{\epsilon^2}, \\
  \hat{P}^2 &\approx -\frac{U(\hat{P},\epsilon)+U(\hat{P},-\epsilon)-2\hat{\mathds{1}}}{\epsilon^2}, \label{eq:X_P_2nd}\\
  \hat{X}^3 &\approx - \frac{i}{2\epsilon^3} \bigl( U(\hat{X},\epsilon)^2-U(\hat{X},-\epsilon)^2 -2 \bigl(U(\hat{X},\epsilon)-U(\hat{X},-\epsilon)\bigr) \bigr), \label{eq:X_3rd} \\
  \hat{X}^4 &\approx \frac{1}{\epsilon^4}\bigl(U(\hat{X},-\epsilon)^2 - 4\,U(\hat{X},-\epsilon) + 6\,\hat{\mathds{1}} - 4\,U(\hat{X},\epsilon) + U(\hat{X},\epsilon)^2\bigr). \label{eq:X_4th}
\end{eqnarray}
For the mixed second-order symmetrized product we use
\begin{eqnarray}
 \hspace{-2cm} \hat{X}\hat{P}+\hat{P}\hat{X} &\approx & -\frac{1}{4\epsilon^2}\Big( U(\hat{X},\epsilon)U(\hat{P},\epsilon) + U(\hat{P},\epsilon)U(\hat{X},\epsilon) - U(\hat{X},\epsilon)U(\hat{P},-\epsilon)  \nonumber \\
 \hspace{-2cm} &&   ~~~~~~~ - U(\hat{P},-\epsilon)U(\hat{X},\epsilon) - U(\hat{X},-\epsilon)U(\hat{P},\epsilon) - U(\hat{P},\epsilon)U(\hat{X},-\epsilon) \nonumber \\ 
 \hspace{-2cm} &&   ~~~~~~~  + U(\hat{X},-\epsilon) U(\hat{P},-\epsilon) + U(\hat{P},-\epsilon)U(\hat{X},-\epsilon) \Big).
  \label{eq:XP_sym}
\end{eqnarray}
Here \(U(\hat{X},\epsilon)^2 \equiv U(\hat{X},\epsilon)U(\hat{X},\epsilon)=U(\hat{X},2\epsilon)\)
(and similarly for \(\hat P\)).

From these relations we see that all requisite powers of \(\hat X\) and \(\hat P\)
may be assembled as linear combinations of the primitive circuits
$\{\hat{\mathds{1}}$, $U(\hat X,\epsilon)$, $U(\hat P,\epsilon)$, $U(\hat X,2\epsilon)$, $U(\hat P,2\epsilon)$, $U(\hat X,\epsilon)U(\hat P,\epsilon)$, $U(\hat P,\epsilon)U(\hat X,\epsilon)\}$. In Fig.~\ref{fig:gateFidelity} we plot the operator gate fidelities
\begin{equation}
F(\epsilon)
  \equiv 1 - \frac{1}{2}\,\bigl\lVert \hat{\mathcal{O}}_\epsilon - \hat{\mathcal{O}} \bigr\rVert_1,
\end{equation}
where $\lVert X \rVert_1$ is the sum of the singular values of $X$ \cite{nielsen2010quantum}.  Here
$\hat{\mathcal{O}}_\epsilon$ denotes the operator obtained from our finite‑displacement unitary
reconstruction at step size~$\epsilon$, and $\hat{\mathcal{O}}$ is the corresponding exact Hermitian
operators (e.g., $\hat X$, $\hat X^2$, $\hat P$, $\{\hat X,\hat P\}$, etc.).  The curves are
shown as functions of $\epsilon$ for simulations performed in a truncated five‑qubit
Hilbert space ($d=2^5=32$) representing a uniform grid over the interval $[0,1)$.  Across
the explored range $10^{-1} \ge \epsilon \ge 10^{-3}$ we observe high fidelities.

\begin{table*}[t]
\centering
\footnotesize
\setlength\tabcolsep{3.5pt}

\begin{tabular}{|c|ccc|ccc|ccc|}
\hline
    & \multicolumn{3}{c|}{$t_{0}=0.0$}
    & \multicolumn{3}{c|}{$t_{0}=0.2$}
    & \multicolumn{3}{c|}{$t_{0}=0.4$}\\ \cline{2-10}
\textbf{Attempt}
    & $\hat{\theta}_{A}=\hat{\mathds{1}}$ & $\hat{\theta}_{A}=\hat{\sigma}_{z}$ & $\mathcal{E}_{C}$
    & $\hat{\theta}_{A}=\hat{\mathds{1}}$ & $\hat{\theta}_{A}=\hat{\sigma}_{z}$ & $\mathcal{E}_{C}$
    & $\hat{\theta}_{A}=\hat{\mathds{1}}$ & $\hat{\theta}_{A}=\hat{\sigma}_{z}$ & $\mathcal{E}_{C}$\\ \hline
1   & 5266 & 5222 &  0.0488 & 6199 & 5094 &  0.1293 & 5102 & 5424 &  0.0526\\ \hline
2   & 6243 & 4605 &  0.0848 & 5727 & 4944 &  0.0671 & 5103 & 5307 &  0.0410\\ \hline
3   & 6560 & 4984 &  0.1544 & 5008 & 5110 &  0.0118 & 4325 & 5928 &  0.0253\\ \hline
4   & 2267 & 5065 & -0.2668 & 5028 & 5841 &  0.0869 & 5296 & 5275 &  0.0571\\ \hline
5   & 5198 & 5001 &  0.0199 & 3140 & 4978 & -0.1882 & 5094 & 5555 &  0.0649\\ \hline
6   & 5533 & 3884 & -0.0583 & 5590 & 4512 &  0.0102 & 4570 & 5193 & -0.0237\\ \hline
7   & 5121 & 5248 &  0.0369 & 5470 & 4953 &  0.0423 & 5537 & 5257 &  0.0794\\ \hline
8   & 5608 & 5171 &  0.0779 & 5050 & 5833 &  0.0883 & 5547 & 4884 &  0.0431\\ \hline
9   & 5320 & 5229 &  0.0549 & 5299 & 5257 &  0.0556 & 5251 & 5406 &  0.0657\\ \hline
10  & 3964 & 5268 & -0.0768 & 5276 & 5335 &  0.0611 & 5056 & 4973 &  0.0029\\ \hline
\textbf{Mean}
    &      &      & \textbf{0.00757}
    &      &      & \textbf{0.03644}
    &      &      & \textbf{0.04083}\\ \hline
\textbf{SEM}
    &      &      & \textbf{0.03714}
    &      &      & \textbf{0.02740}
    &      &      & \textbf{0.00997}\\ \hline
    & \multicolumn{3}{c|}{$t_{0}=0.6$}
    & \multicolumn{3}{c|}{$t_{0}=0.8$}
    & \multicolumn{3}{c|}{$t_{0}=1.0$}\\ \cline{2-10}
\textbf{Attempt}
    & $\hat{\theta}_{A}=\hat{\mathds{1}}$ & $\hat{\theta}_{A}=\hat{\sigma}_{z}$ & $\mathcal{E}_{C}$
    & $\hat{\theta}_{A}=\hat{\mathds{1}}$ & $\hat{\theta}_{A}=\hat{\sigma}_{z}$ & $\mathcal{E}_{C}$
    & $\hat{\theta}_{A}=\hat{\mathds{1}}$ & $\hat{\theta}_{A}=\hat{\sigma}_{z}$ & $\mathcal{E}_{C}$\\ \hline
1   & 4937 & 4925 & -0.0138 & 5207 & 6216 &  0.1423 & 5207 & 5840 &  0.1047\\ \hline
2   & 5133 & 5671 &  0.0804 & 5483 & 5104 &  0.0587 & 6401 & 5823 &  0.2224\\ \hline
3   & 5317 & 5071 &  0.0388 & 4426 & 5008 & -0.0566 & 5868 & 5035 &  0.0903\\ \hline
4   & 4293 & 5992 &  0.0285 & 5972 & 5258 &  0.1230 & 4350 & 5438 & -0.0212\\ \hline
5   & 5615 & 5714 &  0.1329 & 4899 & 6220 &  0.1119 & 5276 & 5110 &  0.0386\\ \hline
6   & 6022 & 5133 &  0.1155 & 5698 & 5682 &  0.1380 & 5086 & 5102 &  0.0188\\ \hline
7   & 5775 & 4378 &  0.0153 & 5066 & 4859 & -0.0075 & 4901 & 4494 & -0.0605\\ \hline
8   & 4943 & 4972 & -0.0085 & 5379 & 5331 &  0.0710 & 5174 & 5347 &  0.0521\\ \hline
9   & 5740 & 4813 &  0.0553 & 6248 & 4765 &  0.1013 & 5142 & 5405 &  0.0547\\ \hline
10  & 5493 & 5437 &  0.0930 & 5272 & 5341 &  0.0613 & 4392 & 5285 & -0.0323\\ \hline
\textbf{Mean}
    &      &      & \textbf{0.05374}
    &      &      & \textbf{0.07434}
    &      &      & \textbf{0.04676}\\ \hline
\textbf{SEM}
    &      &      & \textbf{0.01602}
    &      &      & \textbf{0.02040}
    &      &      & \textbf{0.02569}\\ \hline
\end{tabular}
\caption{IBMQ data for $C(t)$ with $t=\{0,0.2,0.4,0.6,0.8,1.0\}$ and
$\hat{\rho}_{eq}=\ket{0}\bra{0}$. Ten data sets were collected with the
circuits in Fig.~\ref{fig:CZ1} (20000 shots each). Counts in
the $\hat{\theta}_{A}$ columns correspond to measuring $\ket{0}$ on the
control qubit; contributions from $\hat{\rho}_{eq}=\ket{1}\bra{1}$ cancel.}
\label{tab:Crdata}
\end{table*}

\subsection{IBMQ experimental raw data}
Tables \ref{tab:Crdata} and \ref{tab:data} show the raw data from the correlation function and the rate simulations of the metastable spin-$\tfrac{1}{2}$ system, using the \textit{ibm\_brisbane} quantum processor.

\begin{table*}[t]
\centering
\footnotesize
\setlength{\tabcolsep}{3pt} 
\renewcommand{\arraystretch}{1.1} 
\begin{tabular}{|c|ccc|ccc|ccc|}
\hline
 & \multicolumn{3}{c|}{$t_0=0.0$} & \multicolumn{3}{c|}{$t_0=0.2$} & \multicolumn{3}{c|}{$t_0=0.4$} \\ \hline
\textbf{Attempt} & $\hat{\rho}_{\ket{0}\bra{0}}$ & $\hat{\rho}_{\ket{1}\bra{1}}$ & $\mathcal{E}_{H1}$ &$\hat{\rho}_{\ket{0}\bra{0}}$ & $\hat{\rho}_{\ket{1}\bra{1}}$ & $\mathcal{E}_{H1}$ & $\hat{\rho}_{\ket{0}\bra{0}}$ & $\hat{\rho}_{\ket{1}\bra{1}}$ & $\mathcal{E}_{H1}$ \\ \hline
1 & 4816 & 5087 & -0.0009700 & 5218 & 4787 & 0.00005000 & 5339 & 4834 & 0.001730 \\ \hline
2 & 4873 & 5117 & -0.0001000 & 5335 & 4980 & 0.003150 & 5435 & 4929 & 0.003640 \\ \hline
3 & 4840 & 5050 & -0.001100 & 5283 & 4809 & 0.0009200 & 5368 & 4834 & 0.002020 \\ \hline
4 & 4874 & 5083 & -0.0004300 & 4868 & 5166 & 0.0003400 & 5404 & 4866 & 0.002700 \\ \hline
5 & 4863 & 5018 & -0.001190 & 5283 & 4984 & 0.002670 & 5351 & 5032 & 0.003830 \\ \hline
\textbf{Mean} & & & \textbf{-0.0007580} & & & \textbf{0.001426} & & & \textbf{0.002784} \\ \hline
\textbf{SEM} & & & \textbf{0.0002108} & & & \textbf{0.0006264} & & & \textbf{0.0004200} \\ \hline
 & \multicolumn{3}{c|}{$t_0=0.6$} & \multicolumn{3}{c|}{$t_0=0.8$} & \multicolumn{3}{c|}{$t_0=1.0$} \\ \hline
\textbf{Attempt} & $\hat{\rho}_{\ket{0}\bra{0}}$ & $\hat{\rho}_{\ket{1}\bra{1}}$ & $\mathcal{E}_{H1}$ & $\hat{\rho}_{\ket{0}\bra{0}}$ & $\hat{\rho}_{\ket{1}\bra{1}}$ & $\mathcal{E}_{H1}$ & $\hat{\rho}_{eq}=\ket{0}\bra{0}$ & $\hat{\rho}_{eq}=\ket{1}\bra{1}$ & $\mathcal{E}_{H1}$ \\ \hline
1 & 5316 & 4843 & 0.001590 & 5196 & 5020 & 0.002160 & 5573 & 4892 & 0.004650 \\ \hline
2 & 5299 & 5084 & 0.003830 & 5374 & 4941 & 0.003150 & 5413 & 4998 & 0.004110 \\ \hline
3 & 5422 & 4886 & 0.003080 & 5327 & 4962 & 0.002890 & 5420 & 5004 & 0.004240 \\ \hline
4 & 5309 & 4966 & 0.002750 & 5299 & 4918 & 0.002170 & 5314 & 4884 & 0.001980 \\ \hline
5 & 5332 & 4950 & 0.002820 & 5512 & 4962 & 0.004740 & 5157 & 5392 & 0.005490 \\ \hline
\textbf{Mean} & & & \textbf{0.002814} & & & \textbf{0.003022} & & & \textbf{0.004094} \\ \hline
\textbf{SEM} & & & \textbf{0.0003609} & & & \textbf{0.0004719} & & & \textbf{0.0005809} \\ \hline
\end{tabular}
\caption{IBMQ data for $\dot{C}(t)$: For each value of $t= \{ 0, 0.2, 0.4, 0.6, 0.8, 1\}$ and $ \hat{\rho}_{\ket{0}\bra{0}}=\ket{0}\bra{0}$ or $ \hat{\rho}_{\ket{1}\bra{1}}=\ket{1}\bra{1}$, the five sets of the circuits in Fig.~\ref{fig:CH1} were performed and each set is made with 10{,}000 shots. The values in $ \hat{\rho}_{\ket{0}\bra{0}}=\ket{0}\bra{0}$ and $ \hat{\rho}_{\ket{1}\bra{1}}=\ket{1}\bra{1}$ columns represent the counts with measurement outcome $\ket{0}$ in the control qubit. Each pair in these two columns is used to calculate the expectation values $\mathcal{E}_{H1}$.%
\label{tab:data}}
\end{table*}
\end{widetext}


\begin{thebibliography}{}

\bibitem{Schlosshauer2007}
M. A. Schlosshauer, {\em Decoherence: and the Quantum-to-Classical Transition} (Springer, 2007).

\bibitem{Hsieh2013}
C. Hsieh and R. Kapral, Entropy {\bf 16}, 200 (2013).

\bibitem{Hele2017}
T. J. H. Hele, Mol. Phys. {\bf 115}, 1435 (2017).

\bibitem{mcardle2020quantum}
S. McArdle {\em et al.}, Rev. Mod. Phys. {\bf 92}, 015003 (2020).

\bibitem{guo2024experimental}
S. Guo {\em et al.}, Nat. Phys. {\bf 20}, 1 (2024).

\bibitem{chan2023grid}
H. H. S. Chan {\em et al.}, Sci. Adv. {\bf 9}, eabo7484 (2023).

\bibitem{su2021fault}
Y. Su {\em et al.}, PRX Quantum {\bf 2}, 040332 (2021).

\bibitem{bauer2020quantum}
B. Bauer {\em et al.}, Chem. Rev. {\bf 120}, 12685 (2020).

\bibitem{busnaina2024quantum}
J. H. Busnaina {\em et al.}, Nat. Commun. {\bf 15}, 3065 (2024).

\bibitem{chertkov2023characterizing}
E. Chertkov {\em et al.}, Nat. Phys. {\bf 19}, 1799 (2023).

\bibitem{cygorek2022simulation}
M. Cygorek {\em et al.}, Nat. Phys. {\bf 18}, 662 (2022).

\bibitem{bauer2023quantum}
B. Bauer {\em et al.}, PRX Quantum {\bf 4}, 027001 (2023).

\bibitem{davoudi2021toward}
Z. Davoudi, N. M. Linke, and G. Pagano, Phys. Rev. Res. {\bf 3}, 043072 (2021).

\bibitem{schafer2020tools}
F. Sch\"afer  {\em et al.}, Nat. Rev. Phys. {\bf 2}, 411 (2020).

\bibitem{antonini2020cosmology}
S. Antonini and B. Swingle, Nat. Phys. {\bf 16}, 881 (2020).

\bibitem{barata2024probing}
J. Barata and S. Mukherjee, Phys. Rev. D {\bf 111}, L031901 (2025).

\bibitem{viermann2022quantum}
C. Viermann {\em et al.}, Nature {\bf 611}, 260 (2022).

\bibitem{preskill2018quantum}
J. Preskill, Quantum {\bf 2}, 79 (2018).

\bibitem{arute2019quantum}
F. Arute {\em et al.}, Nature {\bf 574}, 505 (2019).

\bibitem{chen2023complexity}
S. Chen {\em et al.}, Nat. Comm. {\bf 14}, 6001 (2023).

\bibitem{samach2022lindblad}
G. O. Samach et al., Phys. Rev. Appl. {\bf 18}, 064056 (2022).

\bibitem{david2023digital}
I. J. David, I. Sinayskiy, and F. Petruccione, Quanta {\bf 12}, 131 (2023).

\bibitem{lee2023evaluating}
S. Lee {\em et al.}, Nat. Comm. {\bf 14}, 1952 (2023).

\bibitem{liu2024simulation}
H. Liu {\em et al.}, Quantum {\bf 9}, 1765 (2025).

\bibitem{guimaraes2024optimized}
J. D. Guimar\~aes {\em et al.}, Phys. Rev. A {\bf 109}, 052224 (2024).

\bibitem{cleve2016efficient}
R. Cleve and C. Wang, {\em the 44th International Colloquium on Automata, Languages, and Programming
(ICALP 2017)} (Schloss Dagstuhl–Leibniz-Zentrum f\"ur Informatik, 2017) p. 17.

\bibitem{joo2023commutation}
J. Joo and T. P. Spiller, New J. Phys. {\bf 25}, 083041 (2023).

\bibitem{campaioli2024quantum}
F. Campaioli, J. H. Cole, and H. Hapuarachchi, PRX Quantum {\bf 5}, 020202 (2024).

\bibitem{CLmodel83}
A. O. Caldeira and A. J. Leggett, Phys. A: Stat. Mech. Appl. {\bf 121}, 587 (1983).

\bibitem{Breuer2002}
H. Breuer and F. Petruccione, {\em The Theory of Open Quantum Systems} (Oxford University Press, Oxford, 2002).

\bibitem{gorini1976}
V. Gorini, A. Kossakowski, and E. C. G. Sudarshan, J. Math. Phys. {\bf 17}, 821 (1976).

\bibitem{lindblad1976generators}
G. Lindblad, Commun. Math. Phys. {\bf 48}, 119 (1976).

\bibitem{mccauley2020accurate}
G. McCauley {\em et al.}, NPJ Quantum Inf. {\bf 6}, 74 (2020).

\bibitem{petruccione}
I. J. David, I. Sinayskiy, and F. Petruccione, Quanta  {\bf 12}, 131 (2023).

\bibitem{manzano2020short}
D. Manzano, AIP Adv. {\bf 10}, 025106 (2020).

\bibitem{johansson2012qutip}
J. R. Johansson, P. D. Nation, and F. Nori, Comput. Phys. Commun. {\bf 183}, 1760 (2012).

\bibitem{yung2012quantum}
M. Yung and A. Aspuru-Guzik, Proc. Natl. Acad. Sci. U.S.A.  {\bf 109}, 754 (2012).

\bibitem{motta2020determining}
M. Motta {\em et al.}, Nat. Phys.  {\bf 16}, 205 (2020).

\bibitem{rall2023thermal}
P. Rall, C. Wang, and P. Wocjan, Quantum  {\bf 7}, 1132 (2023).

\bibitem{childs2017efficient}
A. M. Childs and T. Li, Quantum Inf. Comput. {\bf 17}, 901 (2017).

\bibitem{li2023simulating}
X. Li and C. Wang, {\em the 50th International Colloquium
on Automata, Languages, and Programming (ICALP
2023)} (Schloss Dagstuhl–Leibniz-Zentrum f\"ur Informatik,
2023) p. 87.

\bibitem{chandlerbook}
D. Chandler, {\em Introduction to Modern Statistical Mechanics}
(Oxford University Press, Oxford, 1987).

\bibitem{onsager1931reciprocal}
L. Onsager, Phys. Rev.  {\bf 38}, 2265 (1931). 

\bibitem{berezhkovskii2023population}
A. M. Berezhkovskii and A. Szabo, J. Phys. Chem. B {\bf 127}, 5084 (2023).

\bibitem{craig2004quantum}
I. R. Craig and D. E. Manolopoulos, J. Chem. Phys. {\bf 121}, 3368 (2004).

\bibitem{kabernik2021transition} 
O. Kabernik, Phys. Rev. A {\bf 104}, 052206 (2021).

\bibitem{nielsen2010quantum}
M. A. Nielsen and I. L. Chuang, {\em Quantum Computation and Quantum Information} (Cambridge University Press, 2010).

\end{thebibliography}
\end{document}